\documentclass[conference]{IEEEtran}
\usepackage{moreverb}

\usepackage[table]{xcolor}

\ifCLASSINFOpdf
   \usepackage[pdftex]{graphicx}

   \graphicspath{{../pdf/}{figures/}}
   \DeclareGraphicsExtensions{.pdf,.jpeg,.png,.jpg}
\else
\fi
\usepackage{amsmath}
\usepackage{amssymb}
\usepackage{algorithmic}
\usepackage{algorithm}

\usepackage{array}
\newcolumntype{L}[1]{>{\raggedright\let\newline\\\arraybackslash\hspace{0pt}}m{#1}}
\newcolumntype{C}[1]{>{\centering\let\newline\\\arraybackslash\hspace{0pt}}m{#1}}
\newcolumntype{R}[1]{>{\raggedleft\let\newline\\\arraybackslash\hspace{0pt}}m{#1}}

\def\b0{{\bf 0}}
\def\bb1{{\bf 1}}

\usepackage[colorlinks,bookmarksopen,bookmarksnumbered,citecolor=red,urlcolor=red]{hyperref}

\newcommand\BibTeX{{\rmfamily B\kern-.05em \textsc{i\kern-.025em b}\kern-.08em
T\kern-.1667em\lower.7ex\hbox{E}\kern-.125emX}}

\usepackage{caption}

\usepackage[printonlyused]{acronym}

\acrodef{MAS}{Multi-Agent System}
\acrodef{DC}{Data Center}
\acrodef{VNF}{Virtualized Network Function}
\acrodef{NF}{Network Function}
\acrodef{PAP}{Placement and Assignment Problem}
\acrodef{FPP}{Function Placement Problem}
\acrodef{SPP}{Server Placement Problem}
\acrodef{VDCE}{Virtual Data Center Embedding}
\acrodef{ILP}{Integer Linear Program}
\acrodef{BILP}{Binary Integer Linear Program}
\acrodef{MILP}{Mixed Integer Linear Program}
\acrodef{LARE}{Langragian Relaxation}
\acrodef{VNE}{Virtual Network Embedding}
\acrodef{OTT}{over-the-top}
\acrodef{BPP}{Bin Packing Problem}
\acrodef{MINLP}{Mixed-Integer Nonlinear Programming}
\acrodef{ML}{Machine Learning}
\acrodef{LP}{Linear Program}
\acrodef{CFLP}{Capacitated Facility Location Problem}
\acrodef{ISG}{Industry Standards Group}
\acrodef{ETSI}{European Telecommunications Standards Institute}
\acrodef{PM}{Physical Machine}
\acrodef{VM}{Virtual Machine}
\acrodef{EPC}{Evolved Packet Core}
\acrodef{MCLP}{Maximal Covering Location Problem}
\acrodef{CPRI}{Common Public Radio Interface}
\acrodef{vRAN}{Virtualized Radio Access Network}
\acrodef{LPTRA}{Low Power Transmit and Receive Antenna}
\acrodef{HPTRA}{High Power Transmit and Receive Antenna}
\acrodef{NFV-HRAN}{NFV-based HRAN}
\acrodef{RRH}{Remote Radio Head}
\acrodef{BBP}{Base Band Processing}
\acrodef{BBU}{Base Band Unit}
\acrodef{vBBU}{Virtualized Base Band Unit}
\acrodef{eNodeB}{Evolved Node B}
\acrodef{UE}{User Equipment}
\acrodef{PN}{Physical Network}
\acrodef{VN}{Virtual Network}
\acrodef{HRAN}{Heterogeneous Radio Access Network}
\acrodef{RAN}{Radio Access Network}
\acrodef{CRAN}{Centralized Radio Access Network}
\acrodef{TSP}{Telecommunication Service Provider}
\acrodef{BS}{Base Station}
\acrodef{CSI}{Channel State Information}
\acrodef{QSI}{Queue State Information}
\acrodef{NFV}{Network Function Virtualization}
\acrodef{HPN}{High Power Node}
\acrodef{LPN}{Low Power Node}
\acrodef{NFVI}{Network Function Virtualization Infrastructure}
\acrodef{RL}{Reinforcement Learning}
\acrodef{LTE}{Long Term Evolution}
\acrodefplural{CAPEX}[CAPEX]{CAPital EXpenses}
\acrodefplural{OPEX}[OPEX]{OPerating EXpenses}

\usepackage{algorithmic}
\usepackage{algorithm}
\usepackage{color}
\usepackage[table]{xcolor}
\usepackage{multicol}
\usepackage{multirow}

\usepackage{mdwmath}
\usepackage{mdwtab}
\usepackage{url}

\usepackage{array}
\newcolumntype{L}[1]{>{\raggedright\let\newline\\\arraybackslash\hspace{0pt}}m{#1}}
\newcolumntype{C}[1]{>{\centering\let\newline\\\arraybackslash\hspace{0pt}}m{#1}}
\newcolumntype{R}[1]{>{\raggedleft\let\newline\\\arraybackslash\hspace{0pt}}m{#1}}

\usepackage[tight,footnotesize]{subfigure}
\begin{document}
%
\title{Placement and Scheduling of Functions in Network Function Virtualization}

\author{\IEEEauthorblockN{Rashid Mijumbi, Joan Serrat, JuanLuis Gorricho, Niels Bouten, Filip De Turck, Steven Davy
}
}

\maketitle

\begin{abstract}
The virtualization of \acp{RAN} has been proposed as one of the important use cases of \ac{NFV}. In \acp{VRAN}, some functions from a \ac{BS}, such as those which make up the \ac{BBU}, may be implemented in a shared infrastructure located at either a data center or distributed in network nodes. For the latter option, one challenge is in deciding which subset of the available network nodes can be used to host the physical \ac{BBU} servers (the placement problem), and then to which of the available physical \acp{BBU} each \ac{RRH} should be assigned (the assignment problem). These two problems constitute what we refer to as the \ac{VRAN-PAP}. In this paper, we start by formally defining the \ac{VRAN-PAP} before formulating it as a \ac{BILP} whose objective is to minimize the server and front haul link setup costs as well as the latency between each \ac{RRH} and its assigned \ac{BBU}. Since the \ac{BILP} could become computationally intractable, we also propose a greedy approximation for larger instances of the \ac{VRAN-PAP}. We perform simulations to compare both algorithms in terms of solution quality as well as computation time under varying network sizes and setup budgets.
\end{abstract}


\section{Introduction}
\acp{TSP} currently depend on having proprietary physical devices and equipment for each function making up a given service. As the requirements of users for more diverse and new (short-lived) services increase, \acp{TSP} must correspondingly and continuously purchase, store and operate new physical equipment, which leads to increased \acp{CAPEX} and \acp{OPEX}. Moreover, due to the competition both among themselves and from services provided over-the-top on their data channels, \acp{TSP} cannot respond to this increase in expenses with increased subscriber prices. This has led to dwindling average revenue per user, and forced them to find ways of building more dynamic and agile networks which are cheaper and easier to install and manage so as to remain profitable.\\
\indent \ac{NFV} \cite{nfv, mano} has been proposed as a way to help \acp{TSP} to achieve these goals. The main idea of \ac{NFV} is to decouple \acp{NF} from the physical network equipment on which they run. This can be achieved by taking advantage of recent advances in virtualization technology to pool various \acp{NF} onto high volume servers, switches and storage, which could be located in datacentres, network nodes or user premises. These \acp{VNF} may then be composed and chained to form a service, and could be relocated and instantiated in different network locations without necessarily requiring purchase and installation of new hardware. It is expected that \ac{NFV} will lead to fast, scalable, on-demand and dynamic composition of \acp{NF} to services, and that with these, \acp{TSP} will lower both their CAPEX and OPEX, have more agile networks and enhanced time-to-market.\\
\indent However, since a given service requires a number of VNFs, \ac{NFV} raises two questions; (1) how to define and implement network functions and services, and (2) how to efficiently map and schedule the functions that make up a given service onto a physical network. The European Telecommunications Standards Institute (ETSI) through its NFV industry standards group (ISG) is partnering with network operators and equipment vendors to promote the NFV approach and are currently progressing with regard to the first question above. Specifically, they have already defined the NFV problem, some use cases, a reference architecture and a management and orchestration framework among others \cite{ETSIDOCS}.\\
\indent Nonetheless, the second question i.e. mapping and scheduling has received little attention, with \acp{TSP} and vendors performing this manually in their \ac{NFV} proofs of concept (PoCs). While this may be practical for a PoC involving a single function on a single \ac{VM}, deployments of \ac{NFV} on a large scale will likely present challenges. As networks become bigger and more dynamic, and user service requirements change more often, it will not be possible for \acp{TSP} to manually map and schedule particular VNFs of a given service onto specific physical machines\footnote{This paper uses the terms node and machine synonymously.}. As noted in \cite{Guerzoni12}, automation is paramount to the success of \ac{NFV}. This calls for, among other things, algorithms that are able to perform the mapping of \acp{VNF} onto the possible physical nodes. These algorithms should be able to deal with the online and dynamic nature of services, and must ensure that physical resources are used efficiently. The success of NFV will depend, in part, on the existence and performance of algorithms that determine where, when and how the \acp{VNF} are instantiated \cite{pap}.\\
\indent In this paper, we start by formulating the problem of online mapping and scheduling of VNFs, and then propose algorithms for its solution. In particular, we propose three algorithms that perform the mapping and scheduling of VNFs based on a greedy criterion such as available buffer capacity for the node or the processing time of a given VNF on the possible nodes. The algorithms perform both mapping and scheduling at the same time (one-shot), i.e. at the mapping of each VNF, it is also scheduled for processing. In addition, we propose a local search algorithm based on tabu search (TS) \cite{GloverTS}. The TS algorithm starts by creating an initial solution randomly, which is iteratively improved by searching for better solutions in its neighborhood. Finally, to have a benchmark for comparing performance, we formulate the mapping and scheduling as a \ac{MILP} whose objective is to minimize the time required for packets to be processed by the chain of functions that make up a service. Since a \ac{MILP} could become computationally intractable for practical sizes of the problem, we also propose a heuristic based on hard variable fixing (HVF) \cite{hvfpp}.\\
\indent This paper is a significant extension of our previous work \cite{MijumbiNFV15}. In particular, we extend \cite{MijumbiNFV15} in three major ways: (1) the design and evaluation of the \ac{MILP} and its corresponding HVF heuristic both of which are important in providing a benchmark for evaluations, (2) more evaluations including statistical confidence and computation time, and (3) discussion of the time complexity of the proposed algorithms.\\
%
\indent The rest of this paper is organized as follows: Section \ref{problem} presents a formal description of the mapping and scheduling problem. In Sections \ref{greedyalg} and \ref{tabu}, we describe the proposed greedy and tabu search algorithms respectively. Section \ref{milp} formulates the \ac{MILP} for which a HVF heuristic is proposed in Section \ref{hardvariable}. The algorithms are evaluated in Section \ref{evaluation}, related work discussed in Section \ref{related}, and the paper concluded in Section \ref{concl}.
\section{Problem Description}\label{problem}
\begin{table*}[t]
\begin{minipage}{1\textwidth}
\caption{Variables Associated with the Mapping and Scheduling of Network Functions}
\renewcommand{\arraystretch}{1.3}
\small
\centering
\rowcolors{2}{gray!25}{white}
\begin{tabular}{ c l}\hline
\bfseries Paremeter & \bfseries Description\\
\hline\hline
$1\leq i \leq m$ & subscript for functions of the service\\
$1\leq j \leq n$ & subscript for nodes\\
$\rho_{i,j}$ & processing time of function $i$ on node $j$\\
$\pi_{j}$ & the time of completing the last queued function for processing on node $j$\\
$t_{a}$ & arrival time of a given service\\
$t_{l}$ & deadline for completing the processing of a service\\
$t_{i}$ & completion time of processing function $i$\\
$\delta_{i}$ & buffer requirements for a given function while being processed or queued\\
$B_{j}$ & available buffer resources for a node $j$\\
$\beta_{i,j}$ & eligibility parameter that takes value of 1 if node $j$ is able to process function $i$ and 0 otherwise\\
\hline
\end{tabular}
\label{variables}
\end{minipage}
\end{table*}
The problem of allocating physical resources in NFV can be split into two parts: (1) embedding/mapping virtual machines (VMs) onto physical machines which is known as Virtual Data Center Embedding (VDCE) \cite{Rabbani13}. It is also related to Virtual Network Embedding (VNE) \cite{RashidDissertation, rl, neurofuzzy}. Both VNE and VDCE are well studied and are therefore out of the scope of this paper. The rest of this paper considers that we have \acp{VM} already mapped onto physical machines. (2) mapping and scheduling of VNFs onto the created \acp{VM}. We refer to this problem as network function mapping and scheduling (NFMS), and is the focus of this paper. The NFMS problem results from consideration of the many possibilities for resource sharing in \ac{NFV}. One of them$-$which has been used in most NFV PoCs$-$is that for each VNF, a dedicated VM is used. However, considering a service made up of multiple functions (e.g. a customer premises equipment (CPE) may have about 8 functions \cite{MijumbiNFV15}) means that each customer (or CPE) would require 8 dedicated VMs. This would clearly not be scalable as physical resources would easily be depleted, and would be wasteful of resources since most functions are ``light" and can therefore be processed by a single VM, say, by utilizing software containers and dockers. What we consider in this paper is the resource sharing approach that allows for a given VM to process multiple VNFs, one after another (possibly) from a queue.\\
\indent Therefore, NFMS consists of a need to process network services online (each service is created and embedded as its need arises) using a set $N = \{1, . . . , n\}$ of $n$ virtual network nodes. $N$ represents all the virtual nodes created/mapped on all the physical nodes. Any given network service $S$ is made up of a sequence $F = \{1, . . . , m\}$ of $m$ VNFs, where the function $1\leq i \leq m$ must be processed on a set $N(i) \subseteq N$ of nodes. The functions $\{1, . . . , m\}$ must be processed one after the other in the specified sequence, and each virtual node can process at most one function at a time. The processing time for function $i$ on node $j \in N(i)$ is $\rho_{ij} > 0$, where $1\leq j\leq n$, and while being processed or in the queue for processing, a function $i$ utilizes a buffer $\delta_{i}$ from the node onto which it is mapped. The different function processing times on each node may also capture function setup times, a given function may require different setup times on different nodes. At any point, a given node $j$ has available buffer size $B_j$. For each node-function combination, we define a binary variable $\beta_{i,j}$ which takes a value of 1 if node $j$ is able to process function $i$ and 0 otherwise. We also define a deadline $t_l$ for processing a given service. The processing of the last function in the service must be completed by this time, or the service request is rejected. The deadline can be used to define processing priority for services, e.g. those service that require real time processing may have a deadline that only takes into consideration their processing and precedence requirements on the different virtual nodes, and zero waiting. Finally, for each virtual node $j$, we define the expected completion time $\pi_{j}$ of the last function queued for processing on the node, and for each service, we define an arrival time $t_a$, which is the time when the request for mapping and scheduling the service is received by the physical network.\\
\indent The problem then involves choosing for each VNF $i$ a virtual node $j \in N(i)$ and determining a completion time $t_i$ when its processing will be completed\footnote{The time $t_s$ at which the processing of the function $i$ on node $j$ starts can then be derived from $t_s = t_i - \rho_{ij}$.}. It can be considered to consist of two parts: deciding onto which virtual nodes each VNF should be mapped (the mapping problem), and for each node, deciding the order in which the mapped VNFs should be processed (the scheduling problem.). It is in general possible to solve these two problem in two separate steps, such that VNFs are first mapped to nodes and then scheduled. For ease of reference, Table \ref{variables} summarizes all the variables that define the problem. We will refer to them throughout this paper.

\subsection{Mapping and Scheduling Example}
\subsubsection{Function Mapping}
\begin{figure*}[t]
\setlength{\abovecaptionskip}{7pt plus 0pt minus 0pt}
\setlength{\belowcaptionskip}{-5pt plus 0pt minus 0pt}
\begin{minipage}{.33\textwidth}
\centering
\resizebox{.99\textwidth}{!}
{\includegraphics{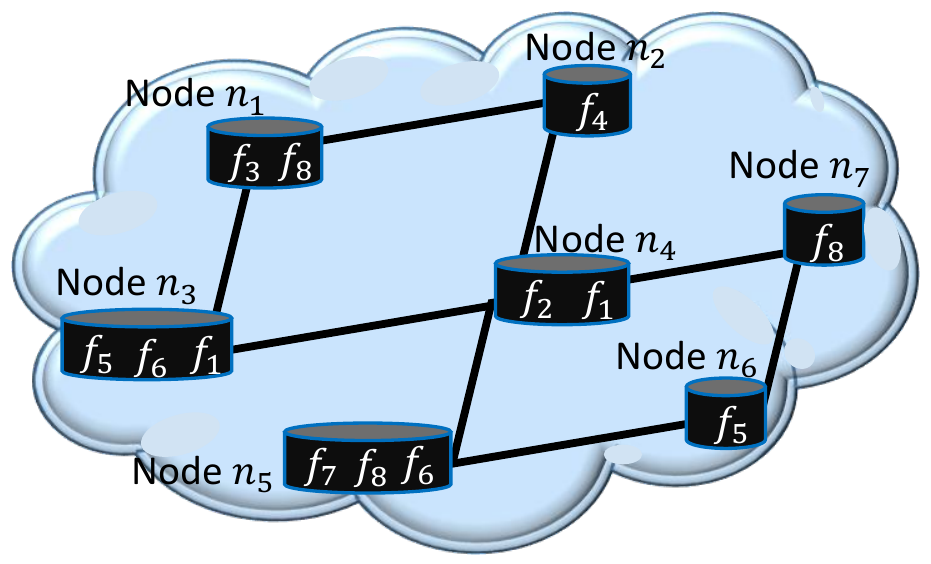}}
  \caption{\scriptsize Network node capabilities}
  \label{state1}
\end{minipage}
\begin{minipage}{.33\textwidth}
\centering
\resizebox{0.99\textwidth}{!}
{\includegraphics{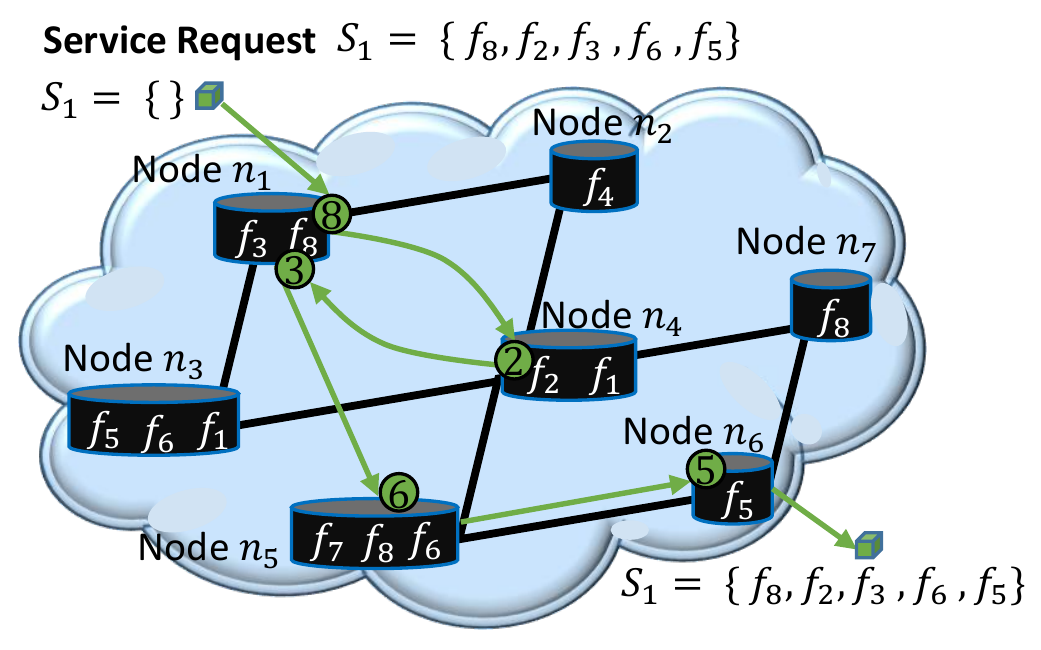}}
  \caption{\scriptsize After mapping of service 1}
  \label{state2}
\end{minipage}
\begin{minipage}{.33\textwidth}
\centering
\resizebox{0.99\textwidth}{!}
{\includegraphics{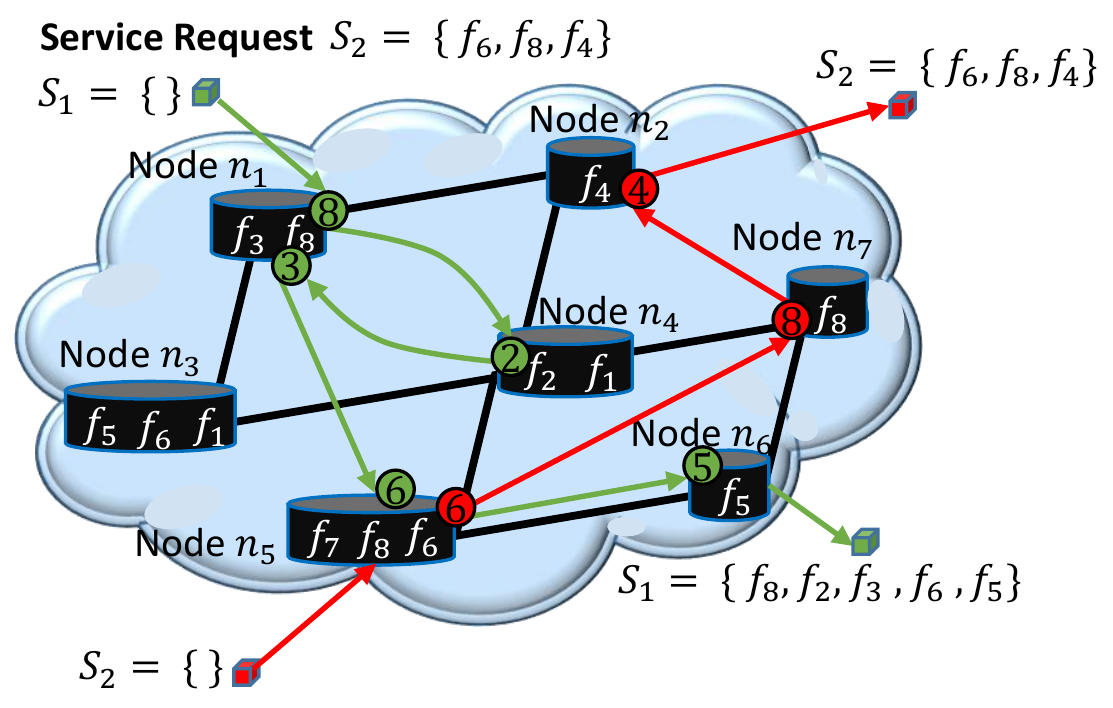}}
  \caption{\scriptsize After mapping of service 2}
  \label{state3}
\end{minipage}
\end{figure*}

As an example, in Figs. \ref{state1} - \ref{scheduling}, we show two services $S_1$ and $S_2$, being mapped and scheduled on a virtual network. In Fig. \ref{state1}, the network is represented showing the different VNF processing capabilities of each node. Consider that at a time $T_1$ a service $S_1$ arrives with the request for mapping with functions $\{f_8\rightarrow f_2\rightarrow f_3\rightarrow f_6\rightarrow f_5\}$. At this point, it is possible that $S_1$ would be mapped as shown in Fig. \ref{state2}. It can be observed that in such a mapping the node $n_1$ is hosting the functions $f_8$ and $f_3$. Similarly, assuming that after some time another service request $S_2$ with functions $\{f_6\rightarrow f_8\rightarrow f_4\}$ arrives, the mapping shown in Fig. \ref{state3} may be achieved. At this point, two services are simultaneously being processed by the virtual nodes. It can be noted that while the function $f_8$ could be processed at node $n_1$ as well, it may be better to use node $n_7$ for $S_2$ to avoid a long queue at $n_1$.
\subsubsection{Function Scheduling}\label{schedex}
Finally, using hypothetical processing times, in Fig. \ref{scheduling}, we represent the possible scheduling of the function processing at each node. The processing of $S_1$ begins immediately on its arrival at $T_1$ at node $n_1$. There after, each of the successive functions has to wait until its preceding function has been processed before its processing can commence. The processing of $S_1$ ends when its last function has been processed at $T_8$. Therefore, the total processing time (flow time) of $S_1$ would be given by $\Big(T_8 - T_1\Big)$, and is equivalent to the summation of the processing times of the functions at the various nodes.\\
\indent However, for illustration purposes, consider that $S_2$ arrives at $T_4$. As can be observed, since its first service is mapped onto node $n_5$ which already has a scheduled function from the first service, the commencement of the processing of $S_2$ has to wait until the completion of function $f_6$ from $S_1$ at time $T_6$. This delay $(T_6 - T_4)$ in commencing the processing of $S_2$ increases its total processing time, and may ultimately lead to inefficient resource utilization since $S_2$ will occupy the virtual nodes for a longer time than it would have without this kind of delay. We should note that this is a direct consequence of the mapping step, since for example, if we had mapped the function $f_6$ of $S_2$ onto $n_3$ (which also has the capacity to process it), then we would have avoided this extra delay on scheduling. Therefore for efficient resource utilization, the mapping and scheduling steps should have some form of coordination.

\subsection{Function Timing Restrictions}\label{mutualexc}
\begin{figure*}[t]
\setlength{\abovecaptionskip}{7pt plus 0pt minus 0pt}
\setlength{\belowcaptionskip}{-5pt plus 0pt minus 0pt}
\begin{minipage}{.99\textwidth}
\centering
\resizebox{.99\textwidth}{!}
{\includegraphics{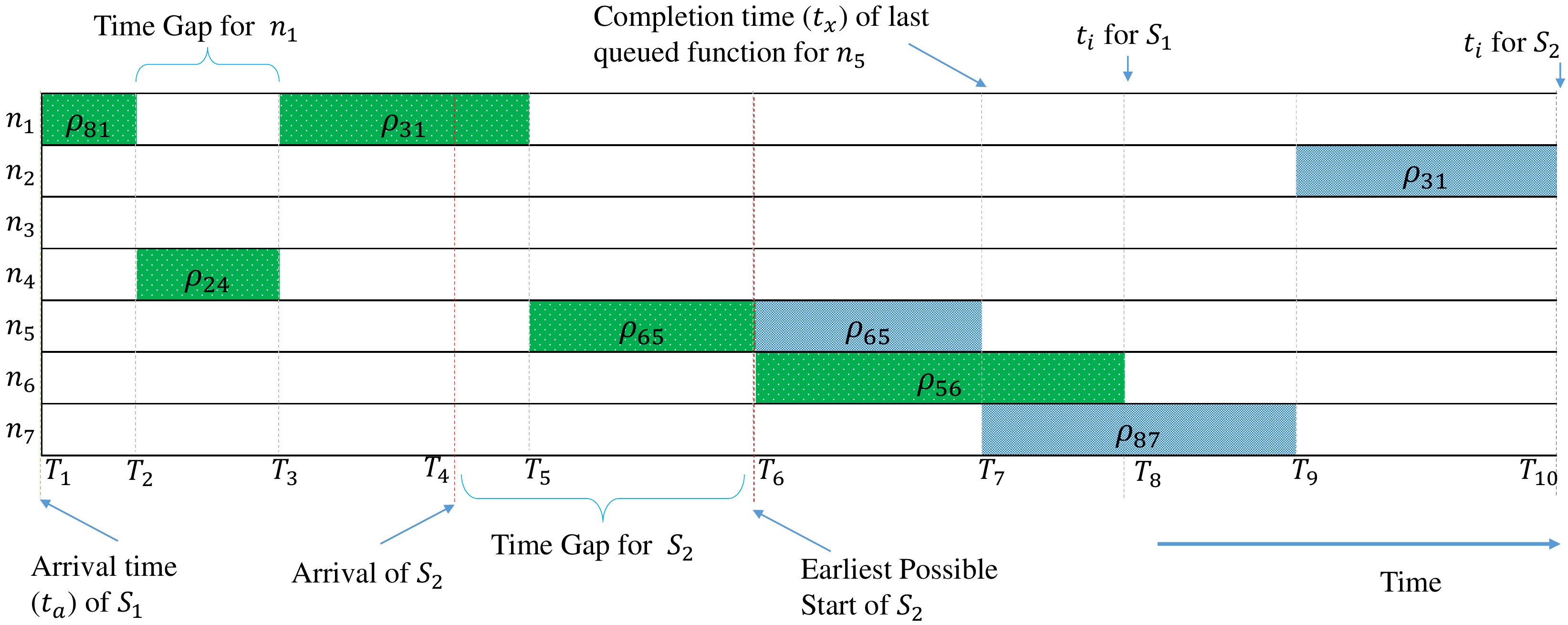}}
  \caption{Function Scheduling}
  \label{scheduling}
  \end{minipage}
\end{figure*}

As illustrated in Section \ref{schedex}, while the processing of a service transits from one function to the next, there may be some delay. There are three mutually exclusive possibilities: (1) immediately after processing a function, the proceeding function is started (for example, in Fig. \ref{scheduling} for $S_1$, the processing of function $f_2$ commences immediately after that of $f_8$), (2) a function (either initial or proceeding) has to wait for the start of its schedule (In Fig. \ref{scheduling}, for $S_2$, assuming that $f_8$ had been mapped onto $n_6$ instead of $n_7$, then after the processing of $f_6$ at time $T_7$, $f_8$ would have had to wait until the node is available.), and (3) the node to which a proceeding function is scheduled has to starve while waiting for the preceding function to be completed (for example, in Fig. \ref{scheduling}, node $n_2$ starves for the time $T_9 - T_4$ waiting to start processing $f_4$. An efficient NFMS algorithm needs to be able to use such timing gaps and delays to ensure both efficient resource utilization and short service processing times.

\subsection{Mapping and Scheduling Objectives}\label{objectives}
In NFMS, there may be many objectives such as minimizing flow time, cost and revenue. 
\subsubsection{Flowtime}
We define the flow time of a service as the difference between when the processing of the last function of a service is completed and when the service arrived. The flow time is a measure of two parameters. On one hand it is a measure of how efficiently resources are being utilized (since a high flow time would mean that a given service occupies the network for extended periods leading to among other things high network loading and hence high power consumption.) while on the other hand it could be used a measure of quality of service if it is associated to the delay of processing a given service. Minimizing flow time means that an average service is processed quickly, at the expense of the largest service taking a long time.
\subsubsection{Revenue}
The revenue $R$ can be defined as the income from the total amount of physical network resources that are utilized by a given mapping and scheduling. It includes the buffer requirements for each function of the service on the node where it is mapped, as well as their processing times. We represent it mathematically in \eqref{rev1}.
\begin{equation}
R = \sum\limits_{i=1}^{m} \delta_{i} + \sum\limits_{i=1}^{m} \sum\limits_{j=1}^{n} \Upsilon_{ij} \times \rho_{ij} 
\label{rev1}
\end{equation}

where $\Upsilon_{ij}$ is a binary variable which is 1 if a function $i$ is mapped on node $j$ and zero otherwise.  
\subsubsection{Cost}
In addition, we can define the cost $C$ as the total amount of physical network resources (both time and buffer) that are utilized by a given mapping and scheduling. In particular, the cost is defined as in equation \eqref{cost1}.

\begin{equation}
C = \theta \sum\limits_{i=1}^{m} \delta_{i} + \varrho \Big(t_i - t_a \Big)
\label{cost1}
\end{equation}

where $\theta$ and $\varrho$ are constants aimed at scaling the costs of buffer and time resources in relation to the revenue. These constants are set as $\theta = \varrho = 0.2$ in this paper. The difference between revenue and cost is in the fact that the revenue only consists of the actual processing times of the functions, while the cost also involves those time gaps that are left unused due to functions waiting for their assigned nodes to become available.
\section{Greedy Function Mapping and Scheduling}\label{greedyalg}
In this Section, we propose three greedy function mapping and scheduling algorithms. The first algorithm, Greedy Fast Processing (GFP), is based on functions being mapped to those nodes that offer the best processing times. The second, Greedy Best Availability (GBA) is based on functions being mapped to those nodes whose current function queue has the earliest completion time. Finally, the Greedy Least Loaded (GLL) algorithm is based on functions being mapped to the node with the highest available buffer capacity. While all these three variations have the potential to minimize the flow time of the mapped service, anyone of them may be used with a specific objective. For example, if services are billed based on the time they spend while being processed, then it may be cheaper to map functions to nodes that process them faster. On the other hand, if the billing is directly proportional the total time the function or service will spend queued for processing or if it is required to perform the processing as early as possible or to balance the  actual loading of the network, then it might be better to use GLL. Balancing the load of the substrate network could lead to a better acceptance ratio for service as it was proved for the embedding of virtual networks in network virtualization environments \cite{Chowdhury12, path}. A combination of these two parameters e.g linear may also be used. Finally, the GBA may be used in case it is required that the service spends the least amount of time in a queue.\\
\indent The algorithms perform as follows: On arrival of a service request, the functions of the service are mapped and scheduled sequentially i.e. the first is mapped and scheduled, then the next one etc. For each function $i$, all the nodes $N(i) \subseteq N$ that have the capacity to process it are determined. These nodes are then ranked based on the greedy criterion, i.e. least loading for GLL, lowest processing times for GFP and shortest virtual node queues for GBA. Then the node with the best rank is chosen for the mapping, and the function is scheduled for processing at the end of the node queue. The actual processing start time is based on both the node being available (completing the processing of previously queued functions), and the the processing of the preceding function (if applicable) being complete. It should also follow all the other constraints as stated in Section \ref{problem}. For example, assuming that the node with the best processing time does not have enough buffer resources for the function being considered, then the second best node is chosen as the one with the best rank. In addition, at each scheduling step, the completion of the processing is determined to ensure that it is within the deadline for the service. The failure of a mapping or scheduling can happen anytime during algorithm run time. For example, it is possible that while mapping the last function of a service, the completion time exceeds the deadline, or that the function has no candidate node (due to all candidate nodes being fully loaded). In this case, the service request is rejected, and all resources allocated to other functions in the service are rolled back. In algorithm 1, we show the psuedocode for these algorithms.
\begin{algorithm}[t]
\caption{Greedy Function Mapping($S$, $N$, $T$)}
\label{greedy}
\begin{algorithmic}[1]
\STATE Backup Substrate Network State
\FOR {Function $i$ $\in$ $S$}
\STATE Initialise: Capable Node Set $N' = \emptyset$
\IF{($ i = 1$)}
\STATE $t_{i-1} = t_a$
\ENDIF
\FOR {Node $j$ $\in$ $N$}	
\STATE $t_e = \rho_{ij} + max (\pi_{j} , t_{i-1})$	
\IF{\Big(($\beta_{ij} == 1$) $\land$ ($B_j$ $\geq$ $\delta_{i}$)$\land$ ($t_e$ $\leq$ $t_{l}$)\Big)}
\STATE $N' = N' \uplus n$
\ENDIF
\ENDFOR
\IF{$N' \equiv \emptyset$}
\STATE NFMS Failed. Reset Resource Status
\STATE \textbf{return}
\ENDIF
\STATE Sort $N'$ according to $T$
\STATE Select the top node $j^*$ from $N'$
\STATE Map the function $i$ onto $j^*$
\STATE Set $t_i = max (\pi_{j} , t_{i-1})$
\STATE Update $B_j$, $\pi_{j}$, and $t_{i-1}$
\ENDFOR
\STATE Mapping and Scheduling Completed
\end{algorithmic}
\end{algorithm}

\subsection{Time Complexity}\label{tc1}
We can now formally analyze the complexity of the greedy algorithms represented by the code in 1. Line 1 involves backing up parameters for each physical network node, and can therefore be done in time linearly proportional to $O(n)$. The for loop from line 7 to 12 includes five operations for each node, and can therefore be done in time $O(5n)$, while lines 13 to 16 can be performed in at most $O(2 + 3n)$. Line 17 involves sorting of a vector of size $n$, which has a average computation complexity of $O(n \log n)$ using the quicksort algorithm \cite{Alsuwaiyel98}. Combining these items, we observe that the for loop from line 2 to 22 can be realized by performing all the above operations for each function. We observe that the dominating factor in the greedy algorithms is the sorting in line 17, which is performed for all the functions. This gives an overall polynomial time computation of $O\big(mn\log n\big)$ for our greedy algorithms.

\section{Tabu Search-based NFMS}\label{tabu}
\subsection{Tabu Search (TS)}
TS is a metaheuristic search method based on local (neighborhood) search methods used for mathematical optimization \cite{Glover1986533}. Local search \cite{MichielsLS} takes a potential solution $Z$ to a problem and checks its immediate neighbors $N(Z)$ in the hope of finding an improved solution $Z'$. The solutions in $N(Z)$ are expected to be similar to $Z$ except for one or two minor details. However, local search methods have a tendency to become stuck in sub-optimal regions or on plateaus where many solutions are equally fit. TS enhances the performance of local search by relaxing its basic rule. First, at each step, worsening moves can be accepted if no improving move is available (like when the search is stuck at a strict local mimimum). In addition, TS uses memory structures that describe the visited solutions or user-provided sets of rules. If a potential solution has been previously visited within a certain short-term period or if it has violated a rule, it is marked as ``tabu" (forbidden) so that the algorithm does not consider moving to that solution repeatedly. A move like this is called a tabu move. However, when a tabu move has a sufficiently attractive evaluation where it would result in a solution better than any visited so far, then its tabu classification may be overridden. A condition that allows such an override to occur is called an aspiration criterion \cite{GloverTS}.

\subsection{Proposed TS Algorithm}
In order to design a TS algorithm, five major components must be determined: the initial solution, the neighborhood solutions, tabu list, aspiration criterion and stopping condition. In what follows, we discuss these aspects with respect the proposed algorithm.
\subsubsection{Initial solution} We start by determining an initial solution $Z_0$. This is determined randomly. It is achieved in two steps: First, for each function $i$ in the service to be mapped, a candidate virtual machine $j$ which meets the requirements described in Section \ref{problem} is randomly chosen from the set $N(i) \subseteq N$. Then, starting with the first one, the functions are scheduled onto the virtual machines where they have been mapped, taking into consideration all the function and machine timing restrictions as described in Section \ref{mutualexc}. The current solution $Z$ is then set as $Z_0$.
\subsubsection{Neighborhood Solutions} In order to find another solution $Z'$ which is better than the current solution $Z$, we need to evaluate solutions $N(Z)$ in the neighborhood of $Z$. To this end, we should first define $N(Z)$. Ideally, all solutions that involve moving each function from one virtual machine to another could produce a different solution. However, this would lead to a big search space. Therefore, we restrict the neighborhood to be based on changes in the mapping of the function with the biggest preceeding time gap. In other words, we evaluate the time gaps between each function. This is the time between the completion of a preceeding function and the start of processing of the current function. The function $f'$ with the biggest time gap is chosen as a candidate for migrating. This way, $N(Z)$ involves all possible solutions which would result from migrating $f'$ from its current virtual node to another virtual node which has the required capabilities to process it. If there is no candidate virtual machine for $f'$, the function with the next biggest time gap is chosen for migration. After the migration, then the scheduling of all the proceeding functions is evaluated to ensure that the flow time is minimum and its according to the restrictions in Section \ref{mutualexc}.

\subsubsection{Tabu List} If a function $i$ has been moved from virtual machine $j_1$ to virtual machine $j_2$, we declare it as tabu to move this function back to $j_1$ during the next $m-1$ iterations. The reason for using $m-1$ is to give a chance for the remaining $m-1$ functions in a service to be moved before the function under consideration can be returned back to its original virtual machine. It is also tabu to choose a solution with a higher flow time than the best known solution. Therefore, the tabu in this paper is recorded in short-term memory as a 2-tupple $T(i, j_1)$.
\subsubsection{Aspiration criterion}We allow for aspiration in which the criterion is to allow a tabu move if it results in a solution with a lower flow time than that of the best known solution $Z^*$, since this would imply that the solution that results from moving $f$ back to $j_1$ has not been previously visited. In addition, if all available moves are classified as tabu due to having a higher flow time than the best known solution, then we determine and select a ``least tabu" move, which is the move with lowest flow time of all the tabu moves.
\subsubsection{Stopping condition} Finally, we have defined two criteria which determine when the algorithm stops: (1) if after $m$ consecutive iterations without an improvement in the flow time, (2) no feasible solution in the neighborhood of solution $Z$ for all functions. Above these, the computation will also be stopped after a maximum number $\kappa$ of total iterations is reached.\\
\indent The psuedocode for the proposed solution is shown in algorithm 2. An initial solution is determined in line 1,  and if we are successful in getting an initial solution, the initialization step in line 6 sets this as the current solution $Z$, and as the best known solution $Z^*$. We also initialize the tabu list $T_l$ as empty. The while loop starting at line 8 will continue searching for a solution until the \textit{stopping condition} is met. In lines 10 to 14, the neighborhood solutions are checked to eliminate those which are tabu. The solution with the lowest flow time is then chosen in Line 15, and the counters in the tabu list are updated in line 17. If the candidate solution has a lower flow time than the current best (line 18), and its features are added to the tabu list (line 19) and it is set as the new best (line 20).
\begin{algorithm}[t]
\caption{Tabu Search-based NFMS($S$, $N$)}
\label{tscode}
\begin{algorithmic}[1]
\STATE Determine Initial Solution $Z_0$
\IF {$Z_0$ notPossible}
\STATE Mapping and Scheduling Failed
\STATE \textbf{return}
\ENDIF
\STATE Initialize: $Z = Z_0$, $Z^* = Z_0$, $T_l = \emptyset$
\STATE Determine function $f$ to move and neighborhood $N(Z)$
\WHILE {StopConditionNotMet()} 
\STATE solutionSet $= \emptyset$ 
\FOR {sCandidate $\in N(Z)$} 
\IF {sCandidate \textbf{doesNotViolate} $T_l$} 
\STATE solutionSet$=$solutionSet $\uplus$ sCandidate
\ENDIF
\ENDFOR
\STATE sCandidate $=$ BestFlowTime(solutionSet)
\STATE $Z =$ sCandidate
\STATE updateTabuList($T_l$)
\IF {FlowTime(sCandidate) $<$ FlowTime($S^*$)} 
\STATE addNewTabu($T_l, f$, getNode($f$), getSize($S$))
\STATE $Z^* =$ sCandidate
\ENDIF
\ENDWHILE
\STATE return $Z^*$
\end{algorithmic}
\end{algorithm}
\subsubsection{Time Complexity}
The time complexity of the proposed TS approach can be analyzed using the code in algorithm 2. The initial solution in Line 1 requires the same steps as the greedy algorithms except the sorting. Therefore, using the analysis given in Section \ref{tc1}, we determine the complexity of Line 1 as $O(5mn)$. Line 7 involves determining the time gaps for each function, which requires time linearly proportional to $O(m)$. The same line also involves sorting the functions to determine the function with the biggest time gap. This requires $O(m \log m)$. To determine a new scheduling of all the mapped functions, we require$-$at most$-$a time of $O(mn)$. Therefore, the total time required for Line 7 is $O\big(m(1+n+ \log m)\big)$. Lines 10 to 14, involve 2 operations on a maximum neighborhood set whose maximum size is the number of nodes, giving a complexity of $O(2n)$. Line 15 sorts all possible solutions which can be done in a maximum time of $O(n \log n)$. Therefore, Lines 9 to 21 can be computed in a total time $O(6 + 2n + n \log n)$. Since the algorithm is capped by the maximum allowed number of iterations $\kappa$, the while loop in Lines 8 to 22 can be computed in time $O\big(\kappa(6 + 2n + n \log n)\big)$. Therefore, the significant computation terms for TS are either Line 7 or Lines 7 - 22. Since both these terms are polynomial, the computation time of the TS algorithm is polynomial time.

\begin{table*}[htb]
\begin{minipage}{1\textwidth}
\caption {Definition of Parameters and Variables for the \ac{MILP} Formulation}
\renewcommand{\arraystretch}{1.7}
\small
\centering
\resizebox{\textwidth}{!}{%
\rowcolors{2}{gray!25}{white}
\begin{tabular}{c l}\hline
\bfseries Variable & \bfseries  Description\\
\hline\hline
$\xi_{i,j,k} \in \{0,1\}$ & binary decision variable taking value 1 if function $i$ of the service is processed on the $k^{th} $ position of node $j$\\
$t_{i} \geq 0$ & continuous decision variable for processing completion time of function $i$  \\
$\tau_{j,k} \geq 0$ & continuous decision variable for finishing time of each processing position $k$ on node $j$\\
$1\leq k \leq k_j$  & subscript for processing positions of node $j$\\
$1\leq p \leq p_j$  & subscript for processors of node $j$\\
\hline
\end{tabular}}
\label{variabley}
\end{minipage}
\end{table*}

\section{Mixed Integer Linear Program (MILP)}\label{milp}

\subsection{Variable Definition}

The mapping and scheduling of a function onto a node involves assigning the function to one of the available positions in one of the possible processors. In Table \ref{variabley}, we define the decision variables as well as parameters used in the formulation. This is in addition to the parameters defined in Table \ref{variables}.
\subsection{Objective function}
The objective of the mathematical program proposed in this article is to find feasible function mapping solutions. The formulation may be aimed at different objectives such makespan, tardiness, earliness, flowtime, etc. While makespan  is the most commonly used objective for related flexible job shop scheduling problems \cite{Roshanaei13}, the fact that we consider only a single service request at a time makes makespan equivalent to flowtime. Therefore, our objective is to minimize the flowtime. Minimizing flowtime ensures that there is an efficient utilization of node resources by reducing the idle times, and to some extent ensures that there is an even distribution of node workloads. This in turn means that the availability of each node is maximized by ensuring that the already mapped functions are completed in the shortest possible time. This avoids delays in completion time of newly arriving functions. If the completion time for the last function $m$ is $t_m$ and the service arrived at a time $t_a$, then the flow time is defined as $(t_m - t_a)$. The objective of the mathematical formulation is shown in \eqref{obj}. 

\begin{equation}
Minimize \quad \big(t_m - t_a \big)
\label{obj}
\end{equation}

\subsection{Constraints}

\subsubsection{Node capacity constraints}
\begin{equation}
\sum\limits_{k=1}^{k_j} \sum\limits_{i=1}^{m} \delta_{i} \times \xi_{i,j,k} \leq B_j \ \ \ \ \   \forall j
\label{bc2}
\end{equation}
\begin{equation}
\sum\limits_{k=1}^{k_j}  \xi_{i,j,k} \leq \beta_{i,j} \ \ \ \ \   \forall i, j
\label{c3}
\end{equation}

Constraint set \eqref{bc2} ensures that the node buffer capacity is not exceeded, while \eqref{c3} ensures that a given function is only processed by a given node if the machine has capabilities to process the function, and that for an eligible node, a function may only be processed in utmost one position. This is aimed at ensuring that node resources are not misused, i.e. not all existing node processing capabilities should be used for mapping a given function(s) only because they are available. Therefore, constraint \eqref{c3} enforces the difference between a totally flexible mapping (in which all nodes have capabilities to process all possible functions i.e. $\beta_{i,j} = 1 \ \ \forall i, j$), and a partially flexible mapping (in which a given function may only be processed by a given set of nodes which could be less than the set of all available nodes.)

\subsubsection{Assignment and sequencing constraints}
\begin{equation}
\sum\limits_{i=1}^{m} \xi_{i,j,k} \leq 1 \ \ \ \ \   \forall j, k
\label{c2}
\end{equation}

\begin{equation}
\sum\limits_{j=1}^n \sum\limits_{k=1}^{k_j} \xi_{i,j,k} = 1 \ \ \ \ \   \forall i
\label{c1}
\end{equation}

The constraint set in \eqref{c2} states that a given processing position of a given node only processes at most one function 
at a time. This is reasonable since a given function may only be processed after processing of its preceding function. It also ensures that a given node processes a single function at a time. Similarly, the constraints in \eqref{c1} ensure that both the assignment and sequencing of functions in a given service are performed in a single step (based on the subscripts $j$ and $k$). The fact that the constraint is an equality ($=$) ensures that each function is assigned to a physical node, and because it is equal to $1$, then this in turn ensures that each function of the service is mapped to utmost one node.

Combined, the constraints in \eqref{bc2}, \eqref{c3}, \eqref{c2} and \eqref{c1} ensure that objective function in \eqref{obj} satisfies the mapping constraints as stated in Section \ref{problem}, i.e. they produce feasible mapping solutions.

\subsubsection{Function precedence constraints}
\begin{equation}
t_{i} - t_{i-1} - \sum\limits_{j=1}^{n} \sum\limits_{k=1}^{k_j}  \Big(\xi_{i,j,k} \times \rho_{i,j}\Big) \geq 0 \ \ \ \ \   \forall i>1
\label{c4}
\end{equation}

\begin{equation}
\tau_{j,k} - \tau_{j,k-1} - \sum\limits_{i=1}^{m}  \Big(\xi_{i,j,k} \times \rho_{i,j}\Big) \geq 0\ \ \ \ \   \forall j\geq1, k>1
\label{c5}
\end{equation}

Constraint set \eqref{c4} enforces the precedence constraints between functions of the same service. It is aimed at ensuring that a given function $i$ may only start its processing after its preceding function $i-1$ has been processed. In the same way,  \eqref{c5}, is aimed at enforcing precedence requirements for the processing positions of a given node. In particular, it states that a processing position of a given function may only become available (so as to start function processing) atleast after the processing of a function in its preceding position.

\subsubsection{Timing constraints}

\begin{equation}
 t_{i} - \sum\limits_{j=1}^{n} \sum\limits_{k=1}^{k_j}  \Big(\xi_{i,j,k} \times (\rho_{i,j}+t_a)\Big) \geq 0 \ \ \ \ \ i = 1
\label{c16}
\end{equation}
\begin{equation}
\tau_{j,k} - \sum\limits_{i=1}^{m}  \Big(\xi_{i,j,k} \times (\rho_{i,j}+ t_a)\Big) \geq 0 \ \ \ \ \ \forall j\geq1, k = 1
\label{c17}
\end{equation}

\begin{equation}
 t_{i} - \sum\limits_{j=1}^{n} \sum\limits_{k=1}^{k_j}  \Big(\xi_{i,j,k} \times (\rho_{i,j}+\pi_j)\Big) \geq 0 \ \ \ \ \ \forall i
\label{comp1}
\end{equation}

\begin{equation}
\tau_{j,k} - \sum\limits_{i=1}^{m}  \Big(\xi_{i,j,k} \times (\rho_{i,j}+ \pi_j)\Big) \geq 0 \ \ \ \ \ \forall j, k
\label{comp}
\end{equation}


\begin{equation}
\tau_{j,k} - t_{i} + M \xi_{i,j,k} \leq M \ \ \ \ \   \forall i,j,k
\label{c6}
\end{equation}
\begin{equation}
 t_{i} - \tau_{j,k} +  M \xi_{i,j,k} \leq M \ \ \ \ \   \forall i,j,k
\label{c7}
\end{equation}

\begin{equation}
 t_{m}  \leq t_l
\label{c73}
\end{equation}

Where $M$ is a large positive number. Combined together, the constraints in \eqref{c16}, \eqref{c17}, \eqref{comp1}, \eqref{comp}, \eqref{c6} and \eqref{c7} ensure that the processing of a given function of a given service only starts after its arrival. They ensure that the processing of any function on any node only commences when both the service and node are ready. In particular, constraints \eqref{c16} and \eqref{c17} ensure that the sequencing of the first function of each service is consistent with the arrival times of services as well as the completion of already queued services on each node. In the same way, the constraints \eqref{comp1} and \eqref{comp} that a given node may only start processing a given function after completing the processing of the functions that have already been mapped. These are important constraint sets for the online mapping and scheduling problem as they ensure a continuous time axis, and they represent an important difference between the offline and online mapping problems. 

In addition, constraints \eqref{c6} and \eqref{c7} ensure the mutual exclusive nature of the three timing possibilities as described in Section \ref{mutualexc}. The constraint sets in \eqref{comp1} and \eqref{comp} ensures that the processing of function on any given node can be initiated at the end of the processing of the already queued functions.

Finally, \eqref{c73} enforces that the last function in the service is fully processing before its processing deadline.

\subsubsection{Varibale domains}

\begin{equation}
t_{i} \geq 0 \ \ \ \ \   \forall i, j, k
\label{c9}
\end{equation}

\begin{equation}
\ \tau_{j,k} \geq 0 \ \ \ \ \   \forall i, j, k
\label{c9-1}
\end{equation}

\begin{equation}
\xi_{i,j,k} \in \{0,1\} \ \ \ \ \   \forall i, j, k
\label{c10}
\end{equation}

Finally, \eqref{c9} and \eqref{c9-1} ensure that the variables $t_{i}$ and $\tau_{j,k}$ are positive continuous while using \eqref{c10} ensures that $\xi_{i,j,k}$ is binary.

\subsection{Multi-Objective MILP}
It is worth noting that we can transform the objective in \eqref{obj} into a multi-objective optimization problem, say, by using a linear combination of two objectives such as minimizing flowtime and ensuring load balancing for node buffer resources. This is shown in \eqref{momilp}, where the second term represents the second objective of balancing the load. The constants $\alpha$ and 
$\lambda$ could then be used to bias the formulation to any one of the objectives, while $\delta$ is a small value aimed at avoiding division by zero.
\begin{equation}
Minimize \quad \Bigg(\alpha (t_m - t_a) + \lambda \sum\limits_{i=1}^{m}\sum\limits_{j=1}^{n}\sum\limits_{k=1}^{k_j} \frac{(\delta_i \times \xi_{i,j,k})}{\delta + B_j} \Bigg)
\label{momilp}
\end{equation}

\section{Hard Variable Fixing-based Heuristic}\label{hardvariable}

Since solving a mixed integer linear program is known to be computationally intractable \cite{Schrijver86}, the mathematical formulation proposed in Section \ref{milp} could become intractable for big instances of the problem. In this Section, we propose a heuristic approach, HVF, that is based on relaxing the MILP and a hard variable fixing \cite{FischettiLodi2003} of the problem variables. The idea behind such an approach is that we can set the mathematical program to quickly abort execution and return a (possibly infeasible) solution. Such kind of a solution is the linear programming relaxation of the original MILP. Then the obtained solution can be be analyzed and rounding techniques are applied to selected non-zero variables to the nearest integer value. If some variables remain undetermined, the process is repeated, but in this case, the variables that were set in previous rounds are fixed. This means that the linear program solved in future iterations is a restricted linear program. Therefore, at each successive iteration, only a smaller problem size is solved until a final solution is obtained.

For the relaxed linear program (LP), all the equations in the MILP still hold, except for the domain constraint  \eqref{c10}. This constraint is changed to \eqref{c10x}.

\begin{equation}
0 \leq \xi_{i,j,k} \leq 1 \ \ \ \ \   \forall i, j, k
\label{c10x}
\end{equation}

Therefore, the proposed heuristic is as follows: On arrival of a service request, LP is solved. This gives (possibly) multiple values of $\xi_{i,j,k}$ for each function. It can be observed that by virtue of the assignment constraint \eqref{c1}, each function of a given service will have a non-zero value for atleast one processing position in all the available nodes. Starting with the first function of the service, the nodes with non-zero $\xi_{i,j,k}$ values are evaluated on two fronts: (1) the value of $\xi_{i,j,k}$, and (2) the completion time $\pi_j$. These two values are combined based on \eqref{rank} to determine a $ranking$ for each node.
\begin{equation}
ranking = \frac{\xi_{i,j,k}}{\pi_j}
\label{rank}
\end{equation}
The reasoning behind the $ranking$ criterion in \eqref{rank} is that since LP is formulated to minimize the flowtime, it would normally assign very low values to values $\xi_{i,j,k}$ which could easily be removed from the solution, or those that have the potential to give a worse solution. Therefore, our ranking ensures that we are preferring those nodes with higher values of $\xi_{i,j,k}$. In addition, we already have some evaluation on how soon a given node can start the processing of any function in the value $\pi_j$. By using this value in the denominator, we prefer nodes with lower values of $\pi_j$. The function is mapped and scheduled to the node with the highest $ranking$. This is then repeated for all the other functions, taking into considerations the constraints defined in Section \ref{problem}. In algorithm \ref{hardv} we show the psuedo code for the proposed algorithm.

\begin{algorithm}[t]
\caption{HVF ($S$, $N$)}
\label{hardv}
\begin{algorithmic}[1]
\STATE Backup Substrate Network State
\STATE Reduced Service $S^* = S$
\FOR {Function $i$ $\in$ $S$}
\STATE Solve LP-HF ($S^*, N$)
\IF{($ i = 1$)}
\STATE $t_{i-1} = t_a$
\ENDIF
\STATE Initialise Node $j^* = null$, $v = 0$
\FOR {Node $j$ $\in$ $N$}	
\STATE $t_e = \rho_{ij} + max (\pi_{j}, t_{i-1})$
\FOR {Position $k$ $\in$ $k_j$}		
\IF{\Big($(B_j$ $\geq$ $\delta_{i})$ $\land$ $(\chi_{i,j,k} > v)$ $\land$ ($t_e$ $\leq$ $t_{l}$)\Big)}
\STATE $j^* = j$
\ENDIF
\ENDFOR
\ENDFOR
\IF{$(j^* = null)$}
\STATE Mapping Failed. Reset Resources Status
\STATE \textbf{return}
\ENDIF
\STATE Map the function $i$ onto $j^*$
\STATE Set $t_i = max (\pi_{j} , t_{i-1})$
\STATE Update $B_j$, $\pi_{j}$, and $t_{i-1}$
\STATE Delete $i$ from $S^*$
\ENDFOR
\end{algorithmic}
\end{algorithm}

\subsection{Time Complexity}

Line 1 of algorithm \ref{hardv} can be performed in time proportional to $O(n)$, while line 2 can be performed in $O(m)$. Line 4 is a linear program which can be solved in polynomial time given by $O(N^{3.5}L^2)$ using an algorithm proposed by \cite{Karmarkar84}, where $N$ is the dimension of the problem (number of variables) and L is the number of bits in the input. For our mathematical formulation, the number of variables $N = (m + nm^2)$, implying that line 4 can be solved in solved in $O\big((m+nm^2)^{3.5} L^2\big)$. The for loop in lines $9-16$ can be performed in time $O(n(4m + 4))$, while lines $17-24$ can be performed in time $O(2+3m)$. Therefore, the dominating computation requirements are on the solution for the linear program, which is solved for each function. This implies that the hard variable fixing heuristic can be performed in polynomial time given by $O\big(n^{3.5}m^{6.5} L^2\big)$.

\section{Evaluations}\label{evaluation}

\subsection{Simulation Environment}

To evaluate the proposed algorithms, we have implemented a discrete event simulator in Java. In these evaluations, the services to be mapped and scheduled arrive one at a time following a poisson distribution. The mathematical programs are solved using CPLEX 12.6 \cite{CPLEX12.6}. Simulations were run on a Windows 8.1 64-bit operating system machine with 16.00GB RAM and an Intel dual core each with 2.40GHz CPU. 

\subsubsection{Simulation Parameters}
We defined 10 different network functions with unique labels $1 - 10$. Any given service is created as an ordered combination of a given number of these functions. The service arrival rate is 1 service in every 3 time units, and any service utilizes a given node resources until it has been processed by the corresponding node. Unless stated otherwise, the main parameters used in these simulations for creating the physical network and services are chosen randomly following a uniform distribution with minimum and maximum values  shown in Table \ref{simulationP}. The choice of uniform distribution and random nature of these parameters is motivated by evaluations normally performed for the related problem of VNE \cite{Fischer13, neural, sdn}. Each simulation involves services arriving to the physical network, being mapped and scheduled (or rejected in case the constraints cannot be met), and departing from each node after being processed. Simulations are carried out for 1,500 service arrivals. The maximum number of iterations $\kappa$ for the TS algorithm is $500$.

\subsection{Compared Algorithms}
Since we are not aware of any online mapping and scheduling proposals for network functions, we have only compared the performance of our proposed algorithms. The codes in Table \ref{table1} are used to represent each of the algorithms.

\begin{table}[t]   
\begin{minipage}{.99\linewidth}
\centering
\renewcommand{\arraystretch}{1.4}
\rowcolors{2}{white}{gray!25}
\caption{\scriptsize Simulation Parameter Ranges}
\begin{tabular}{l c c}
\hline
\bfseries Parameter & \bfseries Minimum & \bfseries Maximum\\
\hline\hline
Number of nodes & 100 & 100\\
Node buffer capacity & 75 & 100\\
Function processed by each node & 1 & 7\\
Function processing times & 15 & 30\\
Function buffer demand & 20 & 30\\
Number of functions per service & 5 & 10\\
Service processing deadline & 5000 & 10000\\
\hline
\label{simulationP}
\end{tabular}
\end{minipage} 
\begin{minipage}{.99\linewidth}
\caption{\scriptsize Evaluated Algorithms}
\centering
\renewcommand{\arraystretch}{1.6}
\rowcolors{2}{gray!25}{white}
\begin{tabular}{c l}
\hline
\bfseries Code & \bfseries Function Mapping and Scheduling Algorithm\\
\hline\hline
GFP  & Greedy mapping with bias towards Fast Processing\\
GLL & Greedy mapping with bias towards Least Loaded\\
GBA & Greedy mapping with bias towards Best Availability\\
TS & Tabu Search-based NFMS\\
HVF & Hard Variable Fixing Heuristic for MILP\\
MILP & MILP with flow time minimization objective\\
\hline
\label{table1}
\end{tabular}
\end{minipage}       
\end{table}

%


\subsection{Simulation Results}
The results of the simulations are shown in Figures \ref{AcceptanceRatio} - \ref{compT95pct}. In what follows we discuss these results.
\subsubsection{Acceptance ratio} 
\begin{figure*}[t]
\setlength{\abovecaptionskip}{7pt plus 0pt minus 0pt}
\setlength{\belowcaptionskip}{7pt plus 0pt minus 0pt}

\begin{minipage}{.49\textwidth}
\centering
\resizebox{.99\textwidth}{!}
{\includegraphics{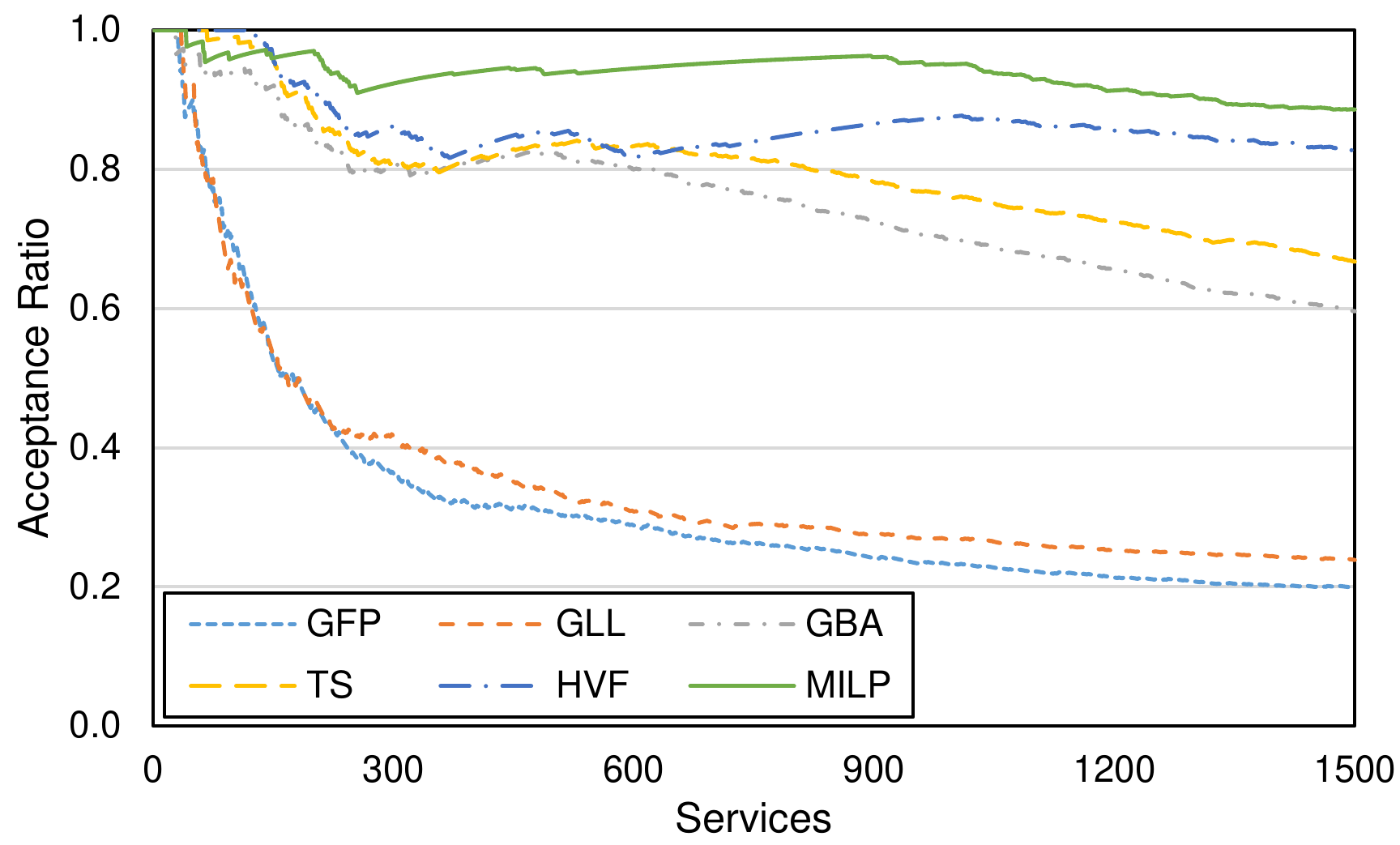}}
  \caption{Service Acceptance Ratio}\label{AcceptanceRatio}
\end{minipage}
\begin{minipage}{.49\textwidth}
\resizebox{.99\textwidth}{!}
{\includegraphics{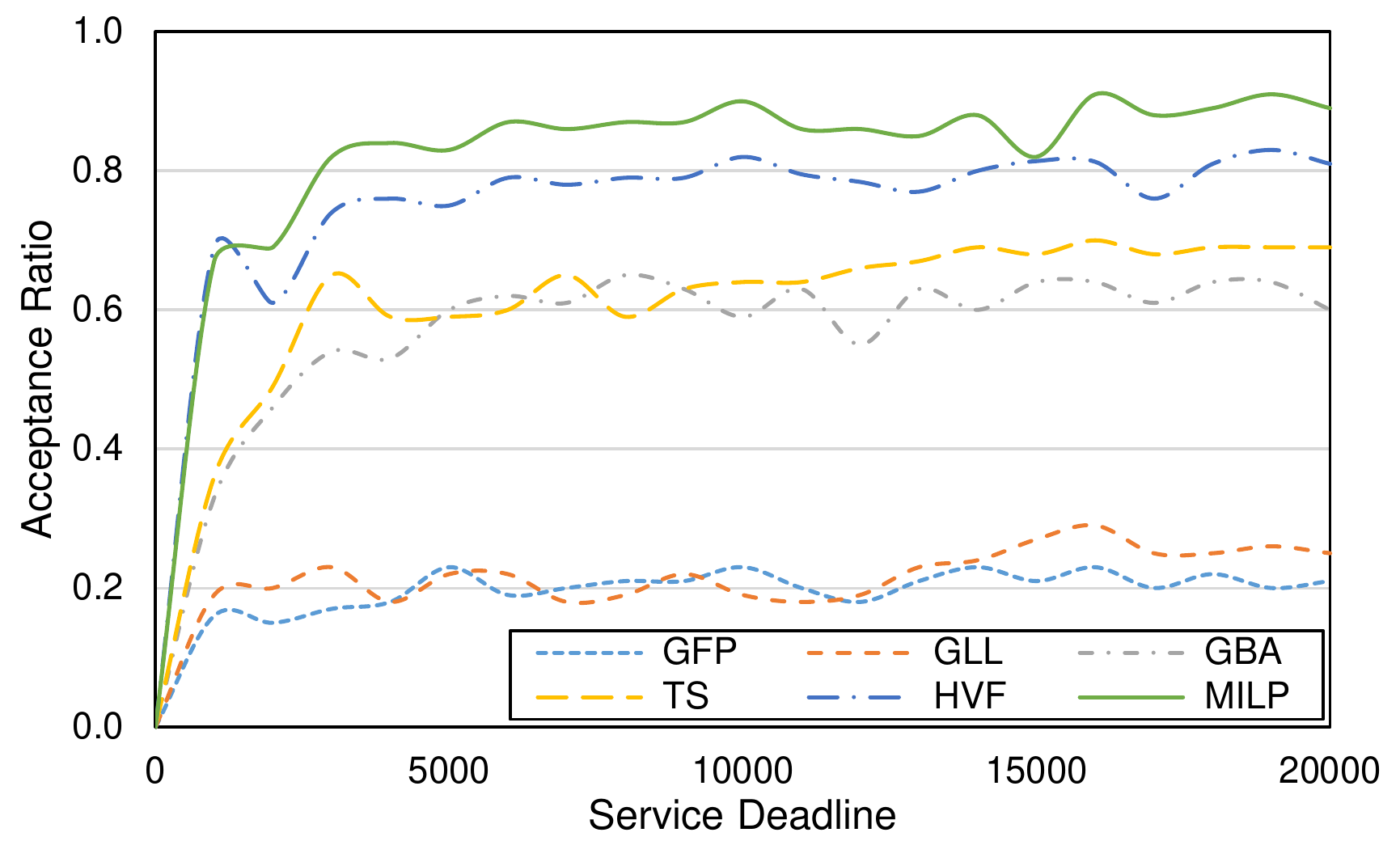}}
  \caption{Effect of Deadline on Acceptance Ratio}\label{effectofD}
\end{minipage}
\end{figure*}

Fig. \ref{AcceptanceRatio} shows the variation of service acceptance ratio with service arrivals. The acceptance ratio is defined as the proportion of the total service requests that are accepted by the network. It is a measure of how efficiently the algorithm uses network resources for accepting service requests\footnote{Once more, this evaluation parameter is motivated by evaluations in VNE.}. As can be noted from the figure, the two algorithms MILP and HVF which are based on mathematical programming perform better than the others. The reason for this is that in the mathematical formulations any given solution (mapping and scheduling) is based on the global knowledge of all the possible solutions. In particular, since the functions to be mapped have restrictions in terms of nodes where they can mapped, performing sequential mapping as is done in greedy algorithms may lead to situations where a function with many mapping options uses up resources of a given node which could possibly be the only option for another function in the same service. It can also be noted that MILP performs better than HVF. This expected, since MILP always finds the optimal mapping and scheduling solution while HVF is a heuristic.\\
\indent In addition, GBA performs significantly better than the other greedy algorithms GFP and GLL, while TS performs better than GBA. The reason why GBA performs well could be attributed to ensuring that services are mapped to nodes which have earlier availability. This ensures that the total time that the service (or any of its functions) takes waiting in a queue for processing is minimized, which does not only avoid holding up resources which could be used by other services, but also ensures that the flow time is within the service requirements. The fact that TS performs better than GBA could be due to the fact that unlike GBA, TS has a chance to iteratively improve the solution. But since the evaluation of new solutions for TS is based on shortest flow time (just like the greedy criterion in GBA), we can observe that the difference in performance for these two algorithms is no so big. Finally, we observe that the greedy approach, which is biased towards favoring using nodes that are least loaded (GLL), performs better than GFP, which is based on favoring nodes with the best processing capacities. This can be attributed to the fact that least loaded nodes are likely to have shorter queues, which implies that services mapped onto such nodes get processed earlier, and hence they do not occupy the node resources for longer periods which could possibly lead to rejecting service requests.
\subsubsection{Effect of Processing Deadline} As defined in the proposed algorithms, each service arrives with a constraint on the latest time that its processing may be completed. In Fig. \ref{effectofD}, we evaluate the effect of this deadline on the acceptance ratio. We vary the service deadline from 0 to 20000 in steps of 1000 and at each step we perform a simulation for 1500 service arrivals. The value of acceptance ratio shown is that after 1500 service arrivals. As can be observed from the figure \ref{effectofD}, the acceptance ratio sharply increases from zero when the deadline is zero before becoming almost constant with increasing deadlines. When service deadlines are very low, they do not even allow for processing of packets at the nodes. As the deadlines increase, nodes are not only able to process some functions, but may also keep some in their queues in cases its required. This accounts for the increasing acceptance ratio early on. At high values of acceptable delays, the acceptance ratio is comparable to the values obtained in Fig. \ref{AcceptanceRatio}, and is not dependent of the delay. This may be attributed to the fact that the networks are allowed to queue up as many services for processing for as long as they have the resources to process them. However, this would be at the expense of completing service processing on time.

\subsubsection{Average Time Gaps and Flow Time} 
\begin{figure*}[t]
\setlength{\abovecaptionskip}{7pt plus 0pt minus 0pt}
\setlength{\belowcaptionskip}{7pt plus 0pt minus 0pt}

\begin{minipage}{.49\textwidth}
\resizebox{.99\textwidth}{!}
{\includegraphics{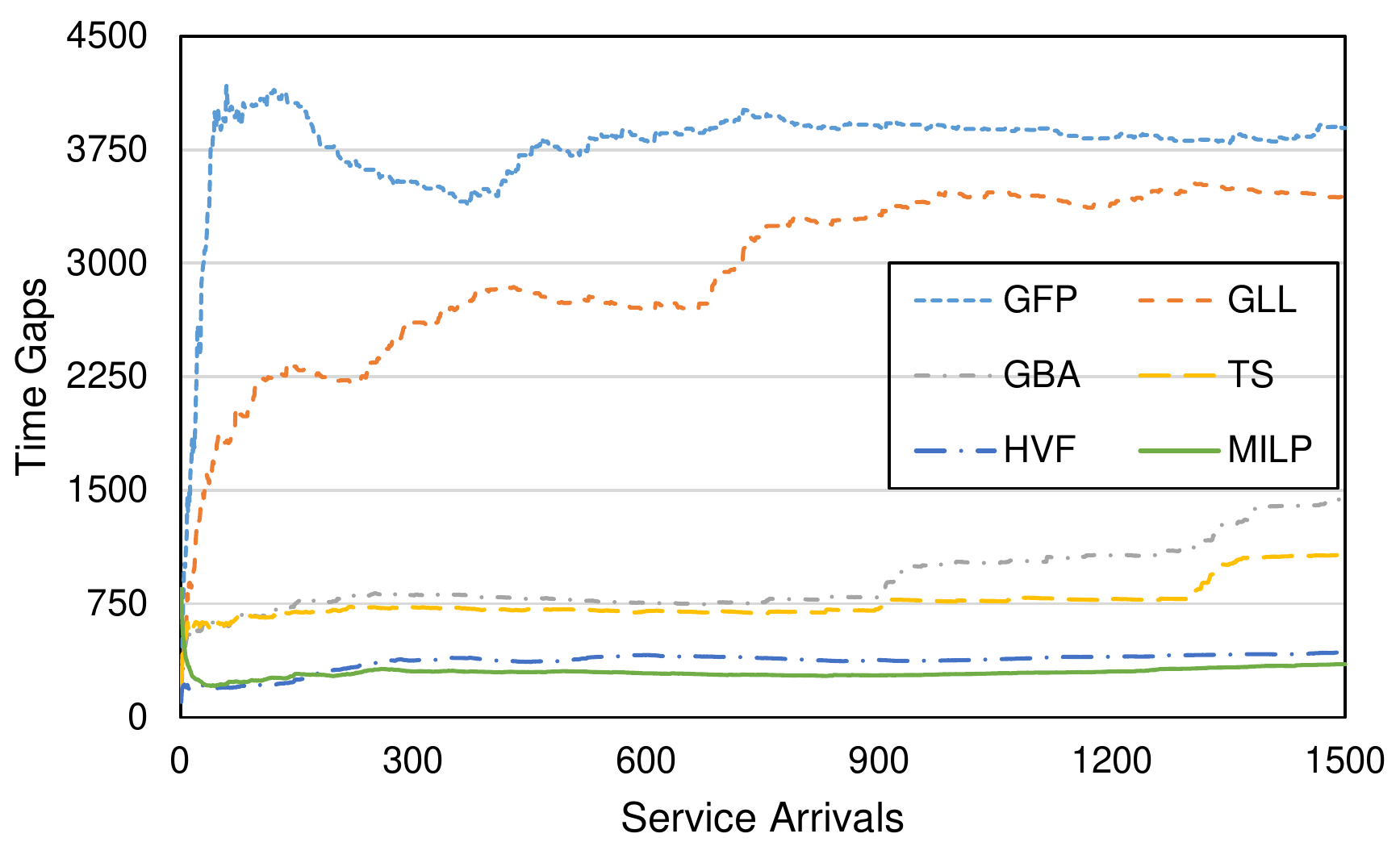}}
  \caption{Time Gaps}\label{timegaps}
\end{minipage}
\begin{minipage}{.49\textwidth}
\resizebox{.99\textwidth}{!}
{\includegraphics{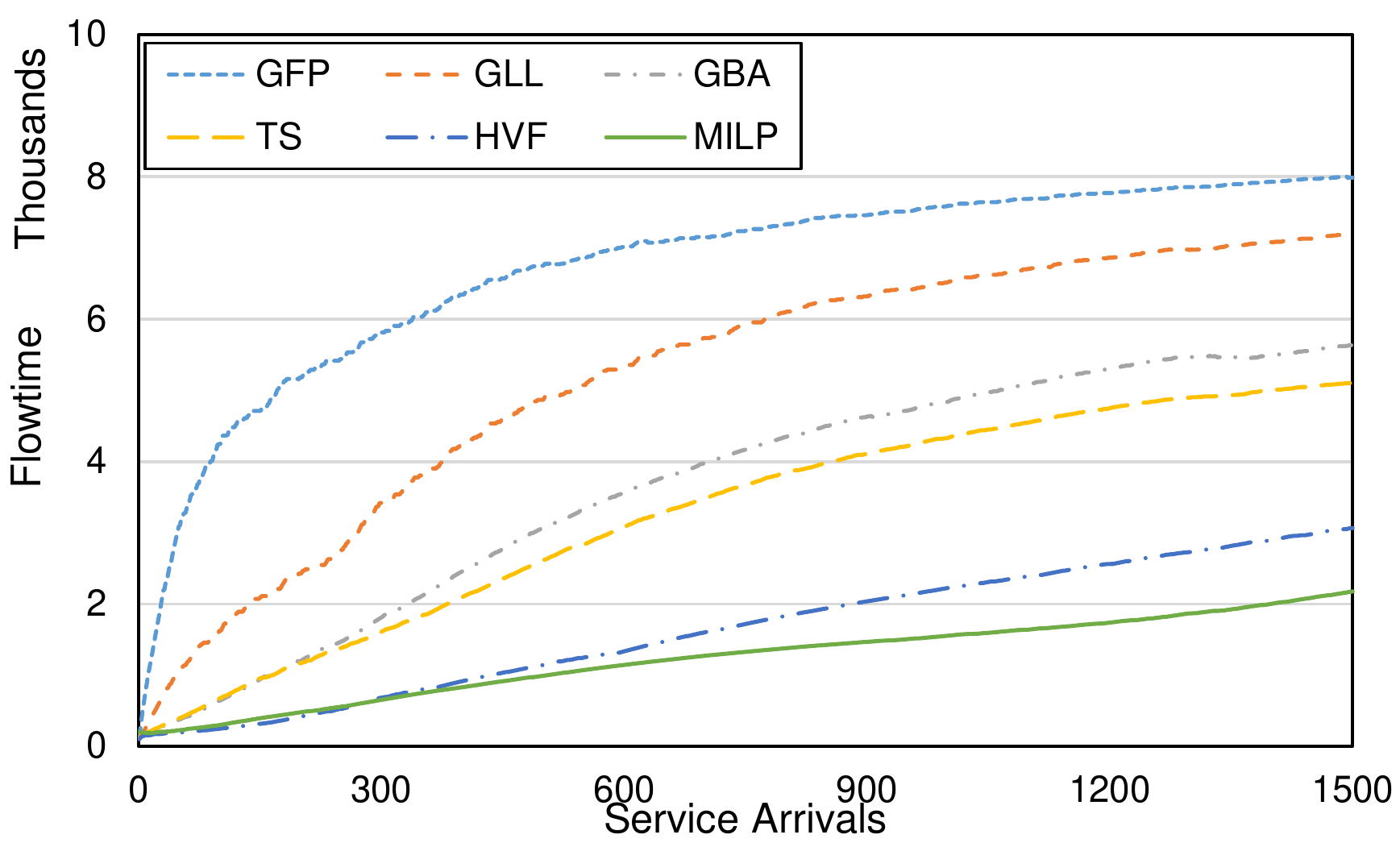}}
  \caption{Total Flow Time}\label{flowtime}
\end{minipage}

\end{figure*}

Figs \ref{timegaps} and \ref{flowtime} show average time gaps and flow time respectively. The time gaps for a given service are determined by summing all those times from when the service arrives when none of its constituent functions is being processed. The average time gaps are determined by averaging the cumulative time gaps of all successfully mapped and scheduled services. Similarly, the flow time is determined as the total time that a given service spends in the network while any of its function is being processed or waiting to be processed. In the same way, values shown in \ref{flowtime} are average values taking into consideration the total successfully mapped and scheduled services. It can be observed that GFP has a considerably higher value of time gaps compared to the other algorithms. This could be attributed to always trying to map a given function to the node that processes it faster. This means that a node that has the least processing time for a given function is likely to always be over loaded, causing a longer queue at such a node, which implies that other functions for the same service should wait for such a function. This is contrary to GLL, for example, which specifically ensures that functions are mapped to those nodes with the lowest loading, which ultimately reduces the waiting times of their executions and hence the time gaps. The reason that TS and GBA perform better than GLL can be attributed to the fact the former algorithms are specifically formulated with the objective of minimizing the flow time. This gives these algorithms an edge in reducing the time gaps, since we could have a scenario where a given node is lightly loaded in terms of buffer/storage utilization, but only because the mapped functions do not require a lot of buffer resources, but take a long time to process. In this way, a function may actually be mapped onto a node where it may have to wait longer, even though this node is not highly loaded. The fact that MILP and HVF have the best performance can be attributed to the mathematical programs being formulated with the objective of minimizing the the flowtime. In addition to their global problem and solution space knowledge advantages, both MILP and HVF may have even more capabilities to further reduce the time gaps, since we could have a scenario where a given node is lightly loaded in terms of buffer utilization, but only because the mapped functions do not require a lot of buffer resources, but take a long time to process. In this way, a function may actually be mapped onto a node where it may have to wait longer, even though this node is not highly loaded. For these reasons, we observe that the actual flow times shown in Fig. \ref{flowtime} have a similar profile as the time gaps in Fig. \ref{timegaps}.

\begin{figure*}[t]
\setlength{\abovecaptionskip}{7pt plus 0pt minus 0pt}
\setlength{\belowcaptionskip}{7pt plus 0pt minus 0pt}

\begin{minipage}{.49\textwidth}
\resizebox{.99\textwidth}{!}
{\includegraphics{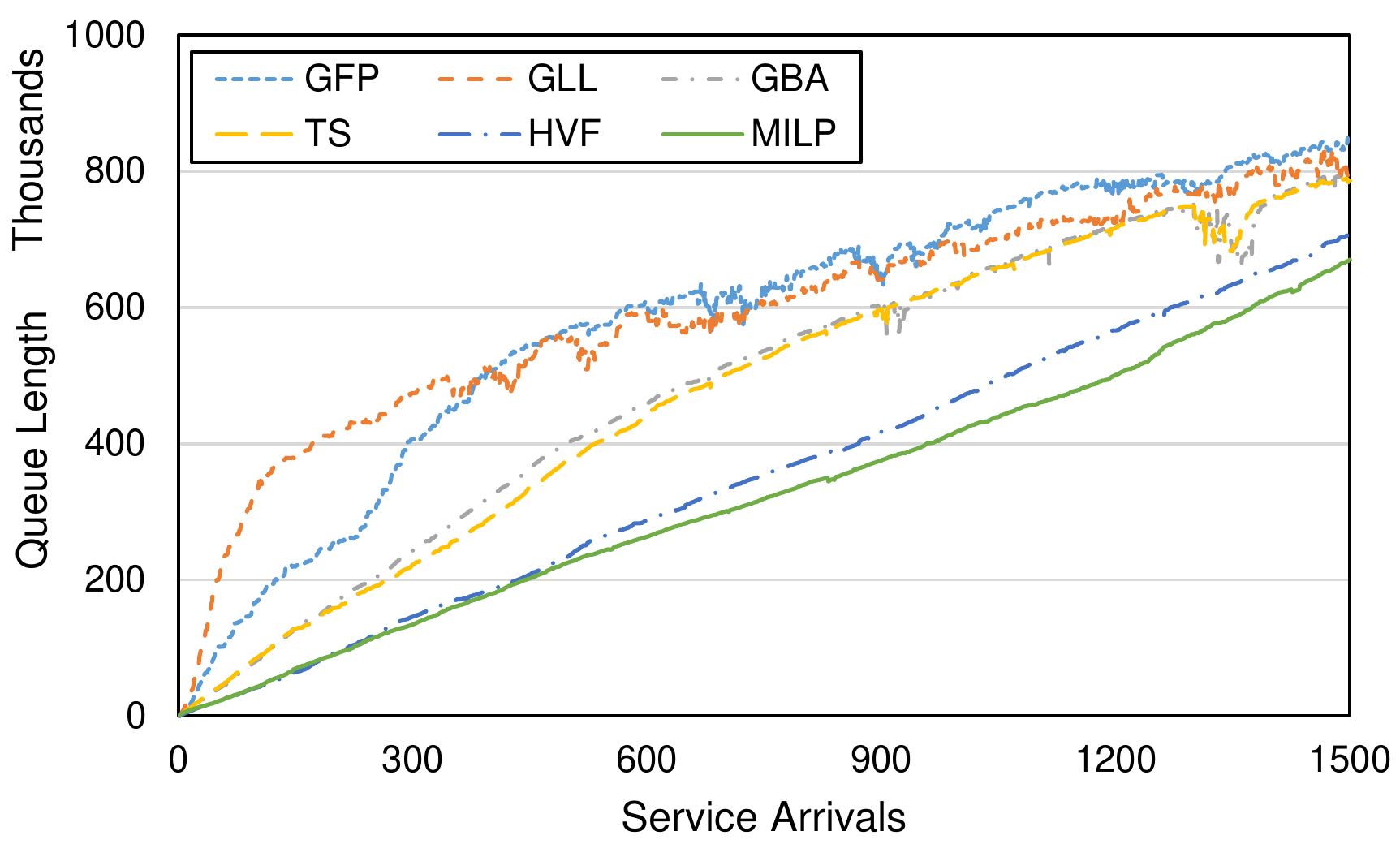}}
  \caption{Queue Length}\label{queuelength}
\end{minipage}
\begin{minipage}{.49\textwidth}
\resizebox{.99\textwidth}{!}
{\includegraphics{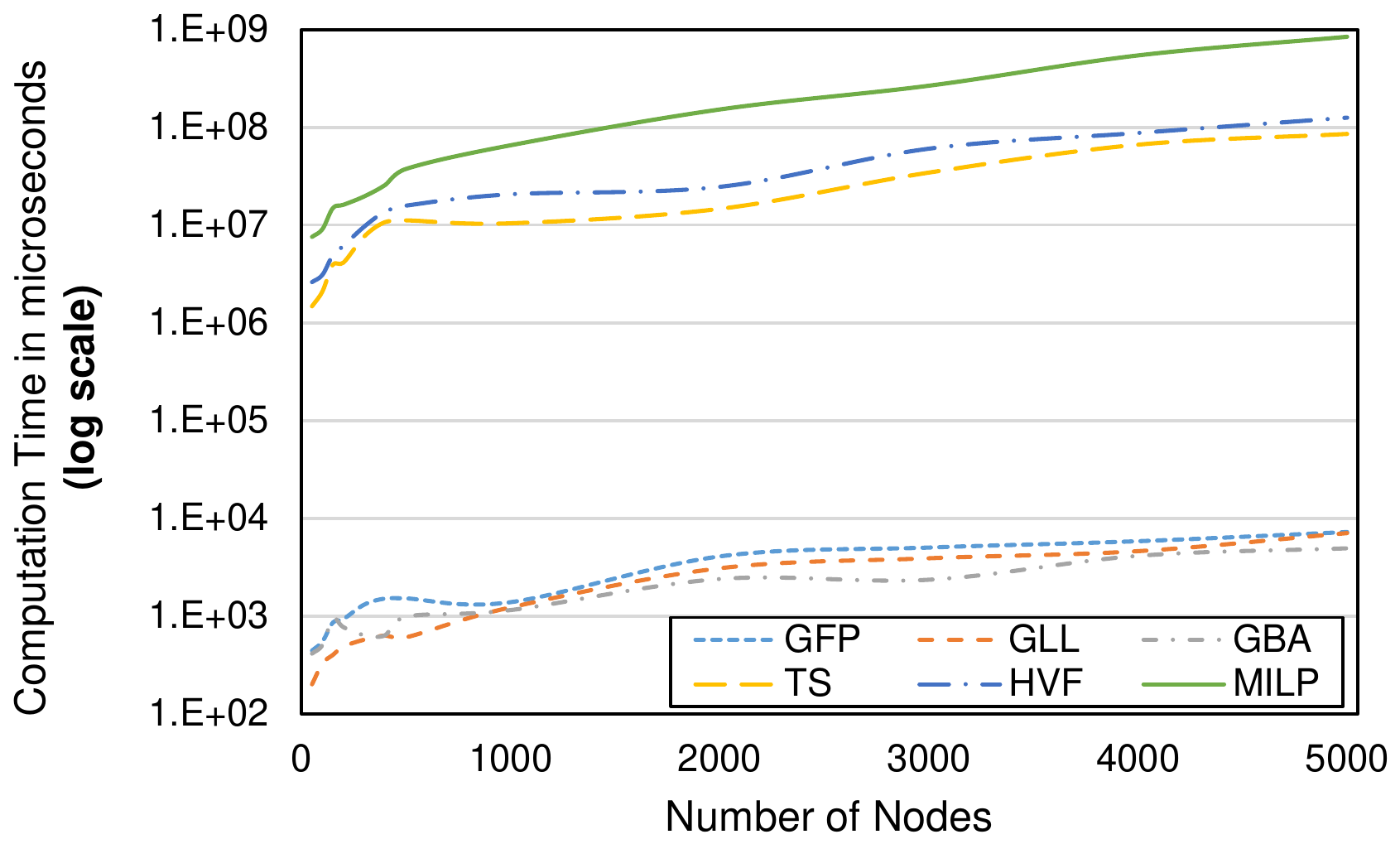}}
  \caption{Computation Time}\label{compt}
\end{minipage}
\end{figure*}

\begin{figure*}[t]
\setlength{\abovecaptionskip}{7pt plus 0pt minus 0pt}
\setlength{\belowcaptionskip}{7pt plus 0pt minus 0pt}
\begin{minipage}{.49\textwidth}
\resizebox{.99\textwidth}{!}
{\includegraphics{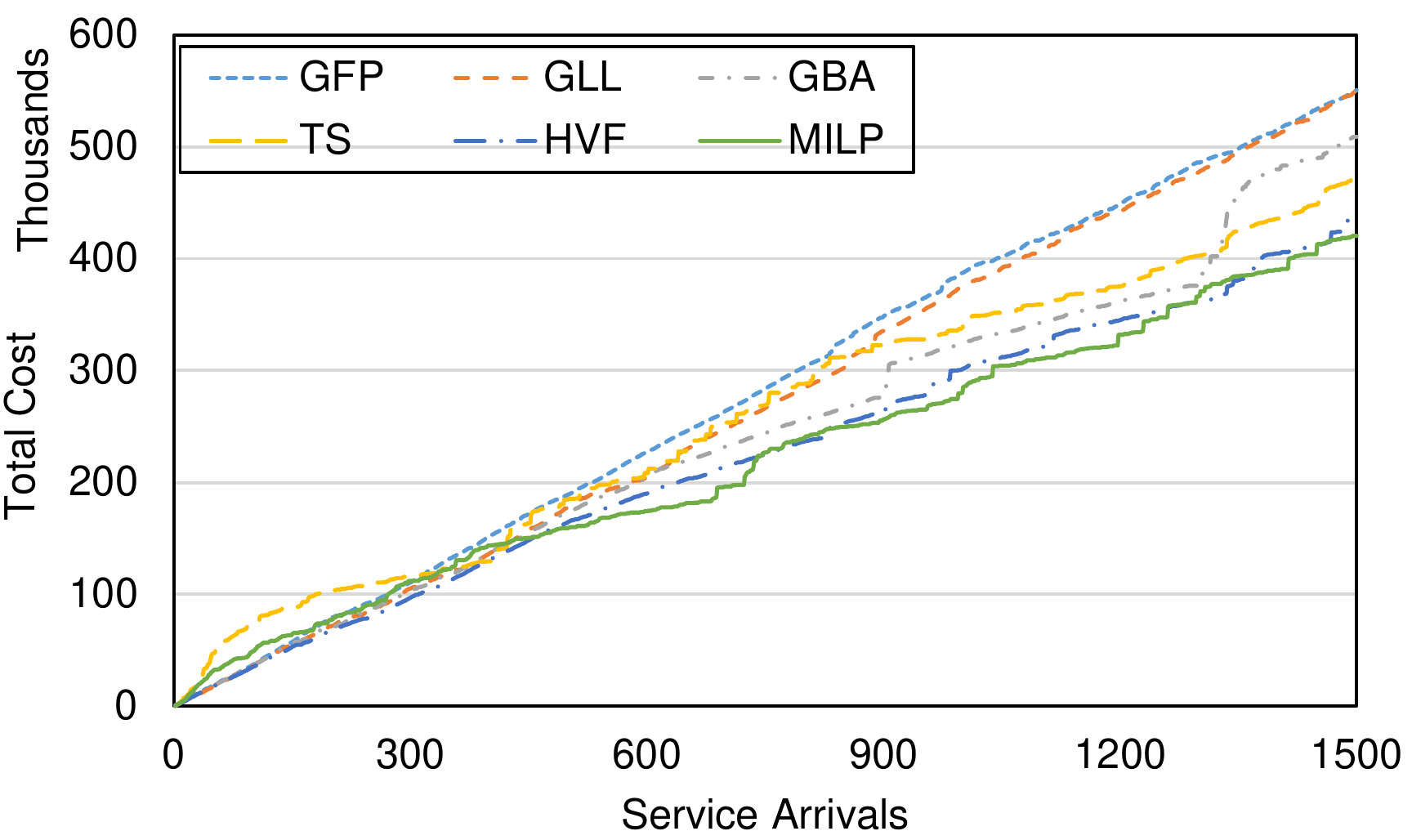}}
  \caption{Total Cost}\label{cost}
\end{minipage}
\begin{minipage}{.49\textwidth}
\resizebox{.99\textwidth}{!}
{\includegraphics{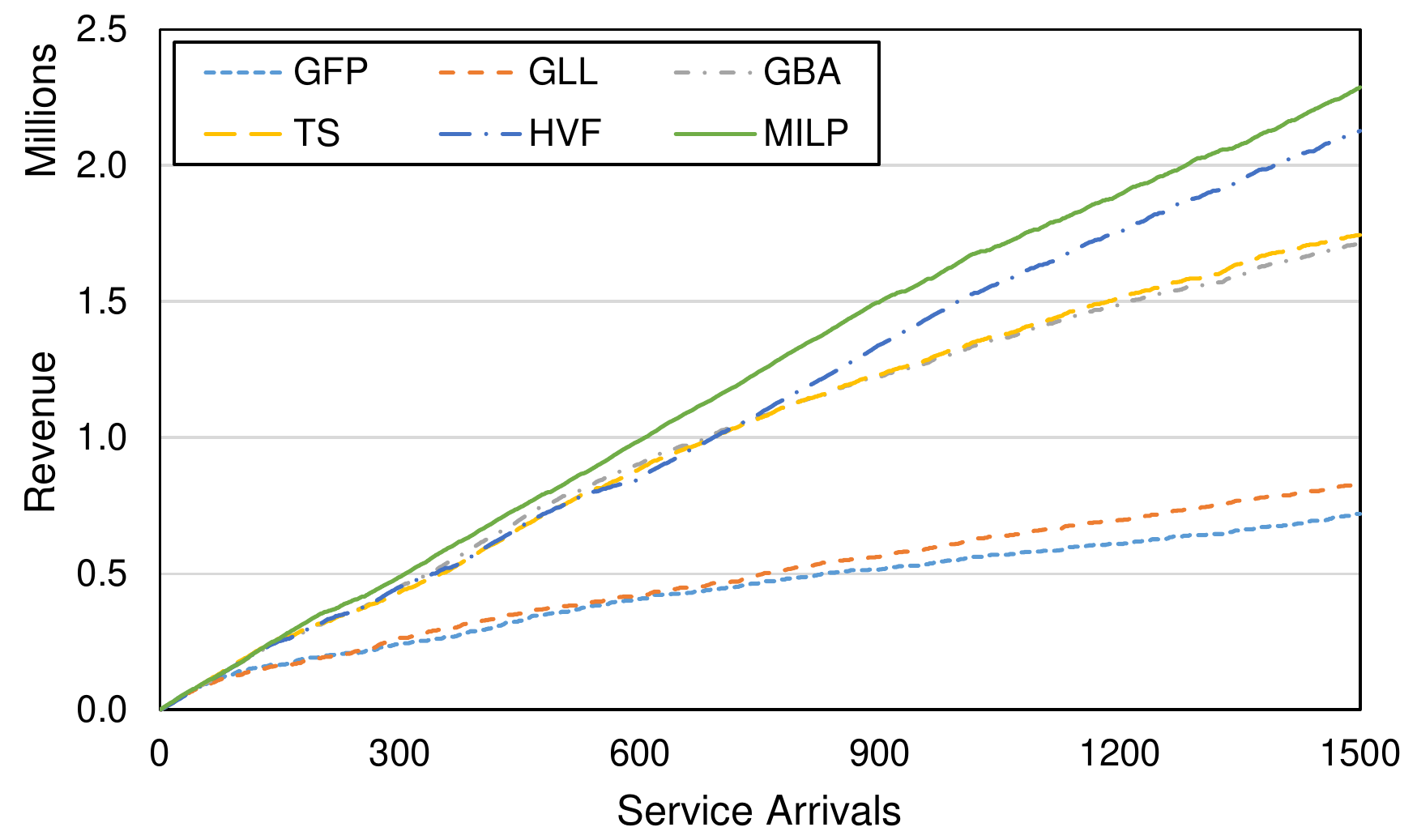}}
  \caption{Total Revenue}\label{revenue}
\end{minipage}
\end{figure*}
\subsubsection{Node Queue Length}
Figure \ref{queuelength} shows the cumulative queue length for all the nodes in the network. We define the queue length of a given network as the total completion time of processing all the functions queued at all its nodes. Subsequent values involve accumulating these values for each accepted service. As expected, we observe that the queue sizes increase as the network accepts more services. In addition, at the beginning, we have a profile similar to that in Fig. \ref{flowtime} where GBA and TS perform better than GFP and GLL, and MILP and HVF better than all other four. Once more this can be explained by the differences in the time gaps from all the algorithms. However, we also observe that as the number of arriving requests increases, the queue of GFP starts increasing at a lower rate, and so does that of GLL. Towards the end of the simulation (1500 service requests), the all the queues are almost comparable, and increase at the same rate. The reason for this could be that at that point, the queues have grown to the maximum sizes that could be permitted by the processing deadlines of each arriving service, hence all services that would make the queues grow higher than these levels are rejected.

\subsubsection{Computation time}In Fig. \ref{compt} we show the computation times (log scale\footnote{We use log scale since the highest computation times are several magnitudes higher than the lowest ones, which would have made the latter invisible on a linear scale.}) for all the algorithms. We define computation time as the time taken to complete the computation of the mapping and scheduling solution using a given algorithm. We observe that the computation times of GFP, GBA and GLL are significantly lower than those of TS, HVF and MILP. This is expected since finding a solution using a mathematical program involves exploring all possible solutions to find the optimal one, as compared to a greedy approach, which only finds a solution on a defined greedy criterion. It can also be observed that the computation time of TS and HVF are much lower than that of MILP. This can be explained by the fact that MILP solves a binary mathematical program while HVF only solves a linear program, which requires less computation resources. In the same way, the computation time of TS is based on multiple individual computations that are polynomial in the size of the physical network and the number of functions. 
\subsubsection{Cost and Revenue}
Figures \ref{cost} and \ref{revenue} show the mapping and scheduling costs as defined in Section \ref{objectives}. The values shown are cumulative, implying that after every successful mapping and scheduling, both the cost and revenues are determined using equations \eqref{cost1} and \eqref{rev1} respectively, and then added to values from the previous accepted service. It can be observed from \ref{cost} that MILP has the lowest cost while GFP the highest, and that these costs increase at almost the same rate for all algorithms. These costs are representative of both the amount of time packets belonging to a given service would require to be processed, as well as the total amount of network resources that would be utilized during that time. It can be observed that, as expected, algorithms that result in higher queues have higher mapping costs. It should however be noted that these costs are not representative of the number of actual functions that are processed by a network at any time. This can be confirmed by the fact that even with the lowest cumulative cost, MILP has the highest acceptance ratio, implying that its average cost per accepted service is much lower. Looking at the cumulative revenue profiles in Fig. \ref{revenue} shows a profile similar to that of the acceptance ratio in Fig. \ref{AcceptanceRatio}, where the more services that are accepted, the higher the revenue. An algorithm that has a higher acceptance ratio is likely to have a high revenue in the long run, which would lead to better profitability for infrastructure providers.

\subsubsection{Statistical bounds}
\begin{figure*}[t]
\setlength{\abovecaptionskip}{7pt plus 0pt minus 0pt}
\setlength{\belowcaptionskip}{7pt plus 0pt minus 0pt}

\begin{minipage}{.49\textwidth}
\resizebox{.99\textwidth}{!}
{\includegraphics{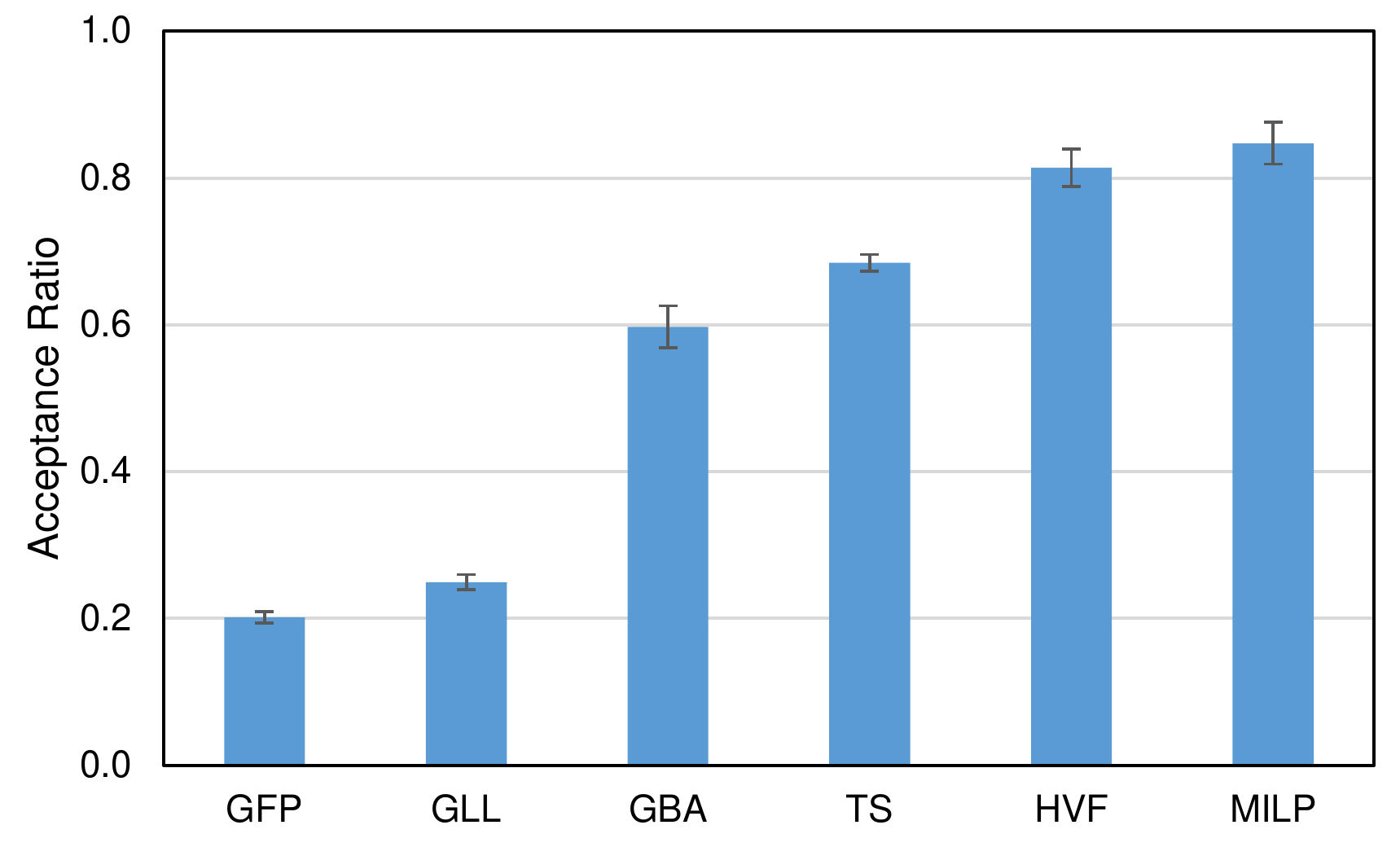}}
  \caption{95\% Confidence on Acceptance Ratio}
  \label{acc95pct}
\end{minipage}
\begin{minipage}{.49\textwidth}
\resizebox{.99\textwidth}{!}
{\includegraphics{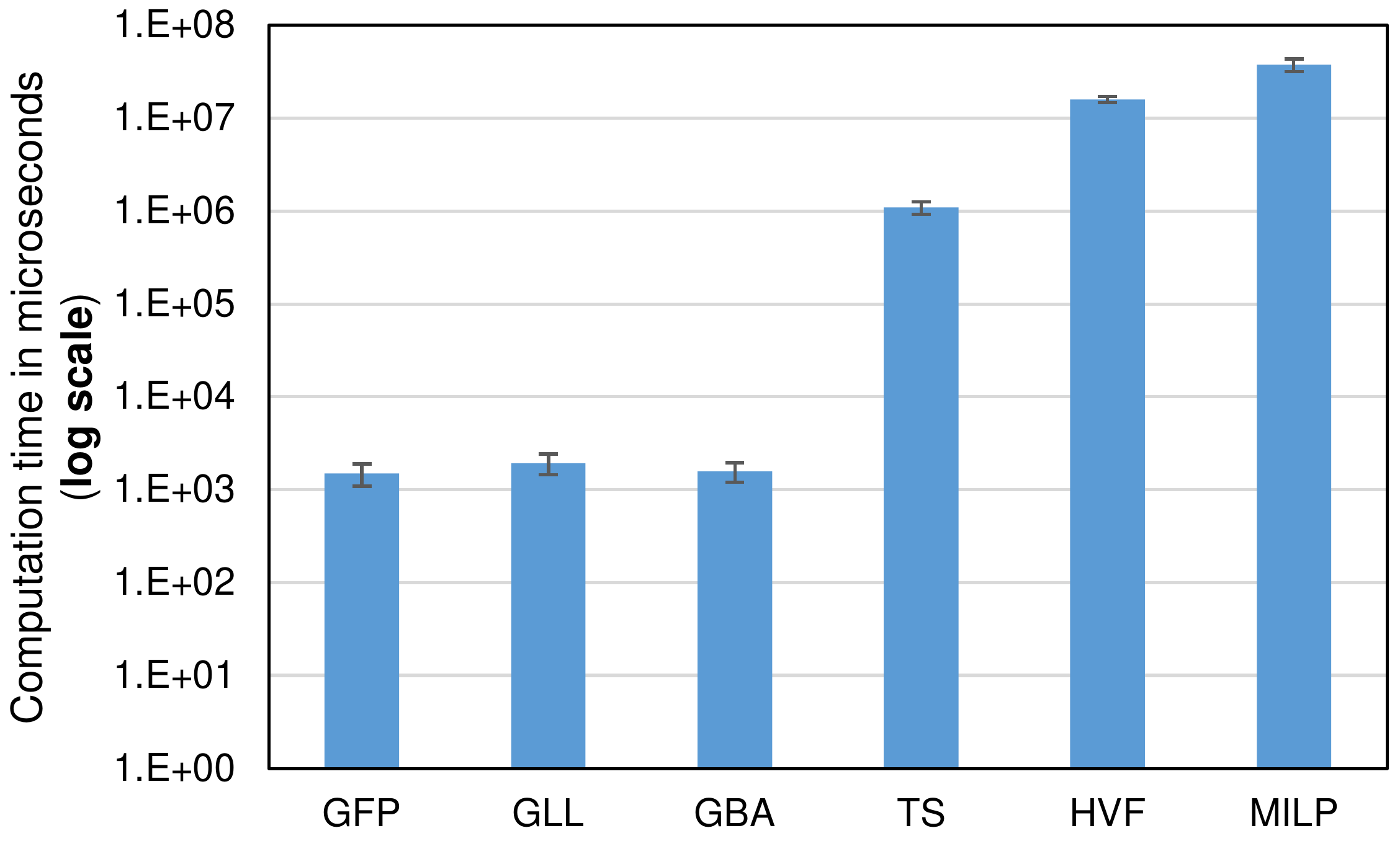}}
  \caption{95\% Confidence on Computation Time}
  \label{compT95pct}
\end{minipage}
\end{figure*}

Finally, in Tables \ref{Tacc95} and \ref{Tacc951}, we show the mean values, standard deviation and 95\% confidence errors margins for the acceptance ratio and computation time. To obtain these values, we repeated the simulations for both the acceptance ratio and computation time 20 times. The final acceptance ratio after 1500 arrivals was obtained, while the computation time is determined for a network of 500 nodes. The values in the tables are represented graphically in Figs. \ref{acc95pct} and \ref{compT95pct} respectively. The small errors at the 95\% confidence level confirm the profiles of the results presented.

\begin{table}[t]
      \caption{Statistical Confidence on Acceptance Ratio}
      \centering
      \renewcommand{\arraystretch}{1.3}
      \rowcolors{2}{gray!25}{white}
\begin{tabular}{l c c c}\hline
 Algorithm &  Mean & Standard Deviation &  95\% Confidence\\
\hline\hline
GFP & 0.20  & 0.02 & $\pm 0.0075$ \\
GLL & 0.25 & 0.02 & $\pm 0.0102$ \\
GBA & 0.60 & 0.07 & $\pm 0.0285$ \\
TS & 0.68 & 0.03 & $\pm 0.0112$ \\
HVF & 0.81 & 0.06 & $\pm 0.0256$ \\
MILP & 0.85 & 0.07 & $\pm 0.0285$ \\
\hline
\label{Tacc95}
\end{tabular}
\end{table}

\begin{table}[t]    
      \centering
      \renewcommand{\arraystretch}{1.3}
      \rowcolors{2}{gray!25}{white}
        \caption{\scriptsize Statistical Confidence on Computation Time}
\begin{tabular}{l c c c}\hline
 Algorithm &  Mean & Standard Deviation &  95\% Confidence\\
\hline\hline
GFP & $1.50\times 10^3$ & $9.21\times 10^2$ & $\pm 4.04\times 10^2$ \\
GLL & $1.94\times 10^3$ & $1.11\times 10^3$ & $\pm 4.88\times 10^2$ \\
GBA & $1.58\times 10^3$ & $8.65\times 10^2$ & $\pm 3.79\times 10^2$ \\
TS & $1.09\times 10^6$ & $3.87\times 10^5$ & $\pm 1.70\times 10^5$ \\
HVF & $1.59\times 10^7$ & $2.67\times 10^6$ & $\pm 1.17\times 10^6$ \\
MILP & $3.76\times 10^7$ & $1.32\times 10^7$ & $\pm 5.80\times 10^6$ \\
\hline
\label{Tacc951}
\end{tabular}
\end{table}

\section{Related Work}\label{related}
\subsection{Virtual Network Embedding}
The mapping of VNFs is related to both VNE \cite{Fischer13, aims} and VDCE \cite{Rabbani13}. In both of these problems, it is required to map virtual resource requests to physical infrastructures in such a way that a given physical node may only map a single virtual node from the same virtual resource request (one-to-one mapping). However, the NFMS problem has a third additional dimension (of mapping network functions) in addition to mapping virtual nodes and links onto a physical network. Moreover, in the mapping of VNFs, a given virtual node may map more than one function from the same service if it is able to process them (possibility of many-to-one mapping). Therefore, the problem involves not only the mapping of virtual machines onto physical machines, but also VNFs onto the virtual machines. It also involves the scheduling of these functions so as to ensure that their timing and precedence requirements are met. VNE does not have this requirement. Finally, as noted by the ETSI ISG on NFV \cite{ETSIDOCS}, ``resource allocation requirements in NFV add new complexity compared to known resource allocation strategies for computing resources in virtualized environments since different VNFs have different latency requirements".

\subsection{Real-time Scheduling}
The function scheduling part of our proposal is related to real-time scheduling \cite{Menglan14}. Real-time scheduling algorithms may be classified as either being offline \cite{JiaXu90, Ramamritham95} or online \cite{Menglan14, Stankovic94}. Static algorithms allocate jobs to processors assuming that all jobs are available to start processing at the same time, while online scheduling is intended for applications in which the jobs or tasks which may unexpectedly arrive \cite{Menglan14}. In this context, we can compare the mapping and scheduling of VNFs to the classical flexible job shop scheduling problem (FJSP) \cite{Chakraborty2009}. The FJSP deals with need to determine a schedule of jobs that have pre-specified operation sequences in a multi-machine environment. While the FJSP is well studied with approaches ranging from meta-heuristics, and artificial intelligence-based approaches \cite{Roshanaei13} almost all approaches in this regard concentrate on the offline problem, in which it is assumed that all the jobs to be scheduled are available at the same time \cite{GuiJuan08}. However, to benefit from the full potential of NFV, it is necessary to allow for an environment where the need to create services appears when there is need. The online scheduling problem is more difficult since we need to consider the arrival times of the requests and there are more possibilities of inefficient resource utilization due to time gaps created by earlier mappings and schedules.

\subsection{NFV Frameworks and Architectures}
There have already been a number of frameworks and architectures proposed for NFV. These include \cite{Guerzoni12} which is a non-proprietary white paper authored by network operators to outline the benefits, enablers and challenges for NFV, \cite{XuJiang14}  an Internet draft that discusses the problem space of NFV and \cite{Boucadair14} a reference architecture and a framework to enforce service function chaining with minimum requirements on the physical topology of the network. In addition, Clayman et al. \cite{SClayman14} describe an architecture that uses an orchestrator to monitor the utilization of infrastructure resources so as to place virtual routers according to policies set by the infrastructure provider, while DROP \cite{Bolla14} proposes a distributed paradigm for NFV through the integration of software defined network \cite{Shid14} and information technology (IT) platforms. T-NOVA \cite{Xilouris14} outlines an architecture, which allows operators to deploy VNFs and offer them to their customers, as value-added services. All these approaches do not address the mapping and scheduling aspects of NFV.
\subsection{Network Function Mapping and Scheduling}
The authors in \cite{moensvnf} present and evaluate a formal model for resource allocation of VNFs within NFV environments, called VNF placement. However, they do not consider the fact that network functions have precedence constraints, hence they consider the problem as a mapping one, similar to VNE. Similarly, \cite{FeRiera14} formulates the problem assuming that nodes have unlimited buffer/storage space to store network functions as they wait for forwarding it to the next node in the sequence. The authors only formulate the problem without solving it. Finally, \cite{SevilM14} proposes an algorithm for finding the placement of the VNFs and chaining them together considering the limited network resources and requirements of the functions. All these proposals consider the offline problem in which all the service requirements are known at the same time.
\section{Conclusion}\label{concl}
In this paper, we have formally defined the problem of mapping and scheduling virtualized network functions onto virtual machines. We have proposed a set of greedy algorithms, a tabu search-based heuristic, a mixed integer linear program and a linear program-based hard variable fixing heuristic. We have extensively evaluated different aspects of these algorithms including acceptance ratio, cost, revenue and computation time and discussed the advantages and disadvantages of each of them. We have also performed a statistical analysis to evaluate the error margins on the presented results. It is our opinion that these algorithms can be used as a starting point for future algorithms and evaluation bench marks. \\
\indent However, the algorithms proposed in this paper do not consider the links between physical nodes, and consequently, the link delays for transferring a given function from one node to another (for processing of the proceeding function) are considered to be negligible. It would be interesting to evaluate the effect of link delays. The algorithms considered in this paper are also static in that after the initial mapping and scheduling, the functions are not migrated with changing network conditions. 

\section*{Acknowledgment}
This work is partly funded by FLAMINGO, a Network of Excellence project (318488) supported by the European Commission under its Seventh Framework Programme, and project TEC2012-38574-C02-02 from Ministerio de Economia y Competitividad.

\bibliographystyle{IEEEtran}
\bibliography{IEEEabrv,nfvbiblio}

\begin{thebibliography}{10}
\providecommand{\url}[1]{#1}
\csname url@samestyle\endcsname
\providecommand{\newblock}{\relax}
\providecommand{\bibinfo}[2]{#2}
\providecommand{\BIBentrySTDinterwordspacing}{\spaceskip=0pt\relax}
\providecommand{\BIBentryALTinterwordstretchfactor}{4}
\providecommand{\BIBentryALTinterwordspacing}{\spaceskip=\fontdimen2\font plus
\BIBentryALTinterwordstretchfactor\fontdimen3\font minus
  \fontdimen4\font\relax}
\providecommand{\BIBforeignlanguage}[2]{{%
\expandafter\ifx\csname l@#1\endcsname\relax
\typeout{** WARNING: IEEEtran.bst: No hyphenation pattern has been}%
\typeout{** loaded for the language `#1'. Using the pattern for}%
\typeout{** the default language instead.}%
\else
\language=\csname l@#1\endcsname
\fi
#2}}
\providecommand{\BIBdecl}{\relax}
\BIBdecl

\bibitem{nfv}
R.~Mijumbi, J.~Serrat, J.-L. Gorricho, N.~Bouten, F.~De~Turck, and R.~Boutaba,
  ``Network function virtualization: State-of-the-art and research
  challenges,'' 2015.

\bibitem{mano}
R.~Mijumbi, J.~Serrat, J.-L. Gorricho, S.~Latr{\'e}, M.~Charalambides, and
  D.~Lopez, ``Management and orchestration challenges in network function
  virtualization,'' 2015.

\bibitem{ETSIDOCS}
{ETSI NFV ISG}, ``{ETSI Network Functions Virtualisation (NFV) Industry
  Standards (ISG) Group Draft Specifications},''
  \url{http://docbox.etsi.org/ISG/NFV/Open}, December 2014, {Accessed on May
  26, 2015}.

\bibitem{Guerzoni12}
R.~Guerzoni, ``{Network Functions Virtualisation: An Introduction, Benefits,
  Enablers, Challenges and Call for Action. Introductory white paper},'' in
  \emph{SDN and OpenFlow World Congress}, June 2012.

\bibitem{pap}
R.~Mijumbi, J.~Serrat, J.-L. Gorricho, N.~Bouten, F.~De~Turck, and S.~Davy,
  ``Design and evaluation of algorithms for mapping and scheduling of virtual
  network functions.''

\bibitem{GloverTS}
F.~Glover and M.~Laguna, \emph{Tabu Search}.\hskip 1em plus 0.5em minus
  0.4em\relax Norwell, MA, USA: Kluwer Academic Publishers, 1997.

\bibitem{hvfpp}
M.~Caserta, S.~Lessmann, and S.~VoB, ``A novel approach to construct discrete
  support vector machine classifiers,'' in \emph{Advances in Data Analysis,
  Data Handling and Business Intelligence}, ser. Studies in Classification,
  Data Analysis, and Knowledge Organization, A.~Fink, B.~Lausen, W.~Seidel, and
  A.~Ultsch, Eds.\hskip 1em plus 0.5em minus 0.4em\relax Springer Berlin
  Heidelberg, 2010, pp. 115--125.

\bibitem{MijumbiNFV15}
R.~Mijumbi, J.~Serrat, J.-L. Gorricho, N.~Bouten, F.~De~Turck, and S.~Davy,
  ``Design and evaluation of algorithms for mapping and scheduling of virtual
  network functions,'' in \emph{IEEE Conference on Network Softwarization
  (NetSoft). University College London}, April 2015.

\bibitem{Rabbani13}
M.~Rabbani, R.~Pereira~Esteves, M.~Podlesny, G.~Simon,
  L.~Zambenedetti~Granville, and R.~Boutaba, ``{On tackling virtual data center
  embedding problem},'' in \emph{{IFIP/IEEE International Symposium on
  Integrated Network Management (IM 2013)}}, May 2013, pp. 177--184.

\bibitem{RashidDissertation}
R.~Mijumbi, J.~Serrat, and J.~L. Gorricho, ``Self-managed resources in network
  virtualisation environments,'' in \emph{IEEE/IFIP International Symposium on
  Integrated Network Management. Ottawa, Canada}, May 2015.

\bibitem{rl}
R.~Mijumbi, J.-L. Gorricho, J.~Serrat, M.~Claeys, F.~De~Turck, and
  S.~Latr{\'e}, ``Design and evaluation of learning algorithms for dynamic
  resource management in virtual networks,'' in \emph{Network Operations and
  Management Symposium (NOMS), 2014 IEEE}.\hskip 1em plus 0.5em minus
  0.4em\relax IEEE, 2014, pp. 1--9.

\bibitem{neurofuzzy}
R.~Mijumbi, J.-L. Gorricho, J.~Serrat, M.~Shen, K.~Xu, and K.~Yang, ``A
  neuro-fuzzy approach to self-management of virtual network resources,''
  \emph{Expert Systems with Applications}, vol.~42, no.~3, pp. 1376--1390,
  2015.

\bibitem{Chowdhury12}
M.~Chowdhury, M.~Rahman, and R.~Boutaba, ``Vineyard: Virtual network embedding
  algorithms with coordinated node and link mapping,'' \emph{Networking,
  IEEE/ACM Transactions on}, vol.~20, no.~1, pp. 206 --219, feb. 2012.

\bibitem{path}
R.~Mijumbi, J.~Serrat, J.-L. Gorricho, and R.~Boutaba, ``A path generation
  approach to embedding of virtual networks,'' \emph{Network and Service
  Management, IEEE Transactions on}, vol.~12, no.~3, pp. 334--348, 2015.

\bibitem{Alsuwaiyel98}
M.~H. Alsuwaiyel, \emph{Algorithms: Design Techniques and Analysis}, ser.
  (Lecture Notes Series on Computing.\hskip 1em plus 0.5em minus 0.4em\relax
  New York, NY, USA: World Scientific Pub Co Inc, 1998, vol.~7.

\bibitem{Glover1986533}
F.~Glover, ``Future paths for integer programming and links to artificial
  intelligence,'' \emph{Computers \& Operations Research}, vol.~13, no.~5, pp.
  533 -- 549, 1986, applications of Integer Programming.

\bibitem{MichielsLS}
W.~Michiels, E.~Aarts, and J.~Korst, \emph{Theoretical Aspects of Local Search
  (Monographs in Theoretical Computer Science. An EATCS Series)}.\hskip 1em
  plus 0.5em minus 0.4em\relax Secaucus, NJ, USA: Springer-Verlag New York,
  Inc., 2007.

\bibitem{Roshanaei13}
V.~Roshanaei, A.~Azab, and H.~ElMaraghy, ``{Mathematical modelling and a
  meta-heuristic for flexible job shop scheduling},'' \emph{International
  Journal of Production Research}, vol.~51, no.~20, pp. 6247--6274, 2013.

\bibitem{Schrijver86}
A.~Schrijver, \emph{Theory of Linear and Integer Programming}.\hskip 1em plus
  0.5em minus 0.4em\relax New York, NY, USA: John Wiley \& Sons, Inc., 1986.

\bibitem{FischettiLodi2003}
M.~Fischetti and A.~Lodi, ``Local branching,'' \emph{Mathematical Programming},
  vol.~98, pp. 23--47, 2003.

\bibitem{Karmarkar84}
N.~Karmarkar, ``A new polynomial-time algorithm for linear programming,''
  \emph{Combinatorica}, vol.~4, no.~4, pp. 373--395, 1984.

\bibitem{CPLEX12.6}
``{IBM ILOG CPLEX Optimizer},''
  \url{http://www-01.ibm.com/software/integration/optimization/cplex-optimizer/about/},
  2014, {Accessed:} 2014-07-13.

\bibitem{Fischer13}
A.~Fischer, J.~Botero, M.~Till~Beck, H.~de~Meer, and X.~Hesselbach, ``Virtual
  network embedding: A survey,'' \emph{Communications Surveys Tutorials, IEEE},
  vol.~15, no.~4, pp. 1888--1906, Fourth 2013.

\bibitem{neural}
R.~Mijumbi, J.-L. Gorricho, J.~Serrat, M.~Claeys, J.~Famaey, and F.~De~Turck,
  ``Neural network-based autonomous allocation of resources in virtual
  networks,'' in \emph{Networks and Communications (EuCNC), 2014 European
  Conference on}.\hskip 1em plus 0.5em minus 0.4em\relax IEEE, 2014, pp. 1--6.

\bibitem{sdn}
R.~Mijumbi, J.~Serrat, J.~Rubio-Loyola, N.~Bouten, F.~De~Turck, and
  S.~Latr{\'e}, ``Dynamic resource management in sdn-based virtualized
  networks,'' in \emph{Network and Service Management (CNSM), 2014 10th
  International Conference on}.\hskip 1em plus 0.5em minus 0.4em\relax IEEE,
  2014, pp. 412--417.

\bibitem{aims}
R.~Mijumbi, J.-L. Gorricho, and J.~Serrat, ``Contributions to efficient
  resource management in virtual networks,'' in \emph{Monitoring and Securing
  Virtualized Networks and Services}.\hskip 1em plus 0.5em minus 0.4em\relax
  Springer Berlin Heidelberg, 2014, pp. 47--51.

\bibitem{Menglan14}
M.~Hu and B.~Veeravalli, ``Dynamic scheduling of hybrid real-time tasks on
  clusters,'' \emph{Computers, IEEE Transactions on}, vol.~63, no.~12, pp.
  2988--2997, Dec 2014.

\bibitem{JiaXu90}
J.~Xu and D.~Parnas, ``Scheduling processes with release times, deadlines,
  precedence and exclusion relations,'' \emph{Software Engineering, IEEE
  Transactions on}, vol.~16, no.~3, pp. 360--369, Mar 1990.

\bibitem{Ramamritham95}
K.~Ramamritham, ``Allocation and scheduling of precedence-related periodic
  tasks,'' \emph{Parallel and Distributed Systems, IEEE Transactions on},
  vol.~6, no.~4, pp. 412--420, Apr 1995.

\bibitem{Stankovic94}
K.~Ramamritham, J.~Stankovic, and P.-F. Shiah, ``Efficient scheduling
  algorithms for real-time multiprocessor systems,'' \emph{Parallel and
  Distributed Systems, IEEE Transactions on}, vol.~1, no.~2, pp. 184--194, Apr
  1990.

\bibitem{Chakraborty2009}
U.~K. Chakraborty, Ed., \emph{Computational Intelligence in Flow Shop and Job
  Shop Scheduling}, ser. Studies in Computational Intelligence.\hskip 1em plus
  0.5em minus 0.4em\relax Berlin: Springer, 2009, vol. 230.

\bibitem{GuiJuan08}
G.-J. Chang, ``On-line job shop scheduling with transfer time in supply
  chain,'' in \emph{Automation and Logistics, 2008. ICAL 2008. IEEE
  International Conference on}, Sept 2008, pp. 284--289.

\bibitem{XuJiang14}
W.~Xu, Y.~Jiang, and C.~Zhou, ``{Problem Statement of Network Functions
  Virtualization Model. Internet-Draft,
  draft-xjz-nfv-model-problem-statement-00},'' Active Internet-Draft, IETF
  Secretariat, Tech. Rep., September 2013.

\bibitem{Boucadair14}
M.~Boucadair, C.~Jacquenet, R.~Parker, D.~Lopez, J.~Guichard, and C.~Pignataro,
  ``{Service Function Chaining: Framework and Architecture. Internet-Draft
  draft-boucadairsfc-framework-02},'' Active Internet-Draft, IETF Secretariat,
  Tech. Rep., February 2014.

\bibitem{SClayman14}
S.~Clayman, E.~Mainiy, A.~Galis, A.~Manzalini, and N.~Mazzocca, ``The dynamic
  placement of virtual network functions,'' in \emph{Proceedings of the
  IEEE/IFIP Network Operations and Management Symposium (NOMS)}, ser. NOMS2014,
  2014, pp. 1--9.

\bibitem{Bolla14}
R.~Bolla, C.~Lombardo, R.~Bruschi, and S.~Mangialardi, ``Dropv2: energy
  efficiency through network function virtualization,'' \emph{Network, IEEE},
  vol.~28, no.~2, pp. 26--32, March 2014.

\bibitem{Shid14}
R.~Mijumbi, J.~Serrat, J.~Rubio-Loyola, N.~Bouten, S.~Latre, and F.~D. Turck,
  ``Dynamic resource management in sdn-based virtualized networks,'' in
  \emph{IEEE International Workshop on Management of SDN and NFV Systems},
  November 2014.

\bibitem{Xilouris14}
G.~Xilouris, E.~Trouva, F.~Lobillo, J.~Soares, J.~Carapinha, M.~McGrath,
  G.~Gardikis, P.~Paglierani, E.~Pallis, L.~Zuccaro, Y.~Rebahi, and A.~Kourtis,
  ``T-nova: A marketplace for virtualized network functions,'' in
  \emph{Networks and Communications (EuCNC), 2014 European Conference on}, June
  2014, pp. 1--5.

\bibitem{moensvnf}
H.~Moens and F.~De~Turck, ``{VNF-P: A Model for Efficient Placement of
  Virtualized Network Functions},'' in \emph{1st IEEE International Workshop on
  Management of SDN and NFV Systems.}, November 2014.

\bibitem{FeRiera14}
J.~Ferrer~Riera, E.~Escalona, J.~Batalle, E.~Grasa, and J.~Garcia-Espin,
  ``Virtual network function scheduling: Concept and challenges,'' in
  \emph{Smart Communications in Network Technologies (SaCoNeT), 2014
  International Conference on}, June 2014, pp. 1--5.

\bibitem{SevilM14}
S.~Mehraghdam, M.~Keller, and H.~Karl, ``Specifying and placing chains of
  virtual network functions,'' in \emph{IEEE 3rd International Conference on
  Cloud Networking (CloudNet)}, Oct 2014, pp. 7--13.

\end{thebibliography}

\end{document}